\documentclass[a4paper,fleqn,usenatbib,useAMS]{mnras}

\usepackage{graphicx}	
\usepackage{amsmath}	
\usepackage{amssymb}	
\usepackage{multicol}        
\usepackage{bm}		
\usepackage{pdflscape}	

\usepackage[T1]{fontenc}
\usepackage{ae,aecompl}

\usepackage{txfonts}

\title[Triggered star formation around bubble N\,24]{Molecular environs and triggered star formation around the large Galactic infrared bubble N\,24}

\author[Li, Esimbek, Zhou, et al.]{Xu Li$^{1,3}$\thanks{Contact e-mail: \href{lixu@xao.ac.cn}{lixu@xao.ac.cn}}
Jarken Esimbek$^{1,2}$\thanks{Contact e-mail: \href{jarken@xao.ac.cn} {jarken@xao.ac.cn}},
Jianjun Zhou$^{1,2}$,
W. A. Baan$^{1,4}$,
\newauthor Weiguang Ji$^{1}$,
Xindi Tang$^{1,2}$,
Gang Wu$^{1,2}$,
Xiaoke Tang$^{1,3}$,
Qiang Li$^{1,3}$,
Yingxiu Ma$^{1}$,
\newauthor Serikbek Sailanbek$^{1,3,5}$,
Dalei Li$^{1,2}$, and Dina Alimbetova$^{5}$\\
\\
$^{1}$Xinjiang Astronomical Observatory, CAS, 150, Science 1-street, Urumqi, Xinjiang 830011, P. R. China\\
$^{2}$Key Laboratory of Radio Astronomy, Chinese Academy of Science, 830011 Urumqi, P. R. China\\
$^{3}$University of Chinese Academy of Science, 19A Yuquan Road, Beijing 100049, P. R. China\\
$^{4}$Netherlands Institute for Radio Astronomy, NL-7991 PD Dwingeloo, the Netherlands\\
$^{5}$Department of Solid State Physics and Nonlinear Physics, Faculty of Physics and Technology, AL-Farabi Kazakh
National University, Almaty 050040, Kazakhstan}

\pubyear{2018}

\begin{document}
\label{firstpage}
\pagerange{\pageref{firstpage}--\pageref{lastpage}}
\maketitle

\begin{abstract}
A multi-wavelength analysis of the large Galactic infrared bubble N\,24 is has been presented in this paper in order to investigate the molecular
and star formation environment around expanding H\,{\sc{ii}} regions. Using archival data from \textit{Herschel} and
ATLASGAL, the distribution and physical properties of the dust over the entire bubble are studied. Twenty three
dense clumps are identified using the Clumpfind2d algorithm with sizes and masses in the range 0.65--1.73 pc and
600--16300 M$_{\odot}$, respectively.
To analyse the molecular environment in N\,24, observations of NH$_{3}$\,(1,1) and (2,2) were carried out using the Nanshan 26-m radio telescope.
Analysis of the kinetic temperature and gravitational stability of these clumps suggests gravitational collapse in several of them.
The mass-size distributions of the clumps and the presence of massive young protostars indicate that the shell of N\,24 is a region of ongoing massive star formation.
The compatibility of the dynamical and fragmentation time-scales and the overabundance of YSOs and clumps on the rim suggest that the "collect and collapse" mechanism is in play at the boundary of the bubble, but the existence of the IRDC at the edge of bubble indicates that "radiation-driven implosion" mechanism may also have played a role there.
\end{abstract}


\section{Introduction}

The infrared (IR) dust bubble, which is a shell structure formed by the interaction of an expanding H\,{\sc{ii}}
region with the surrounding interstellar medium (ISM), provides a target for studying the effects of
massive stellar feedback on the surrounding material.
The last decade has seen great progress towards understanding the nature of bubbles and the star formation
triggered in adjacent shells or bright rimmed clumps \citep[i.e.,][]{deh05, deh06, deh10, zav07, urq07, elm11,
tho12, ken12, ken16, dal15}.
Two mechanisms for triggering star formation around bubbles have been proposed as models: the "collect and
collapse" model \citep[C \& C,][]{elm77} and "radiation-driven implosion" model \citep[RDI,][]{ber89, lef94}.
In the C \& C process the outward expanding  H\,{\sc{ii}} region compress and collect the swept up medium into a region  between the ionization front (IF) and the shock front (SF). This shell between the IF and SF will become denser and may collapse to form stars. By contrast, in the RDI process, the IF drives the SF into the surrounding molecular cloud, stimulating the collapse of pre-existing subcritical clumps to form stars.
Recently, several observational studies and numerical simulations have supported that those two mechanisms,
especially the C \& C mechanism, can successfully explain star formation in several H\,{\sc{ii}} regions,
but they are not conclusive \citep[see i.e.,][]{dal07a, dal07b, lee07, bik10, pet10, bis11, erc12}.
Various efforts have been made to find evidence that bubbles trigger star formation, including individual source studies
\citep[i.e.,][]{ji12, yua14, das16, liu16, dur17} and statistical studies \citep{tho12, ken12}.
 However, the majority of observational studies take a phenomenological approach and frequently conclude with open questions as well as uncertainties
 \citep{ken12}.
 In addition, most of the well studied bubbles are relatively small in size, and investigations of large bubbles
 with radii $>$ 5 arcmin are relatively rare \citep{yua14}.
 A large bubble with a long enough time to sweep up surrounding material would make it easier to find
 evidence of a triggered formation of new generation stars.
 To increase the observational sample of large bubbles for the study of star formation in their vicinity, we have
 selected a large Galactic infrared dust bubble N\,24 cataloged by \citet{chu06} to carry out a multi-wavelength
 and molecular spectral line study.

\begin{figure*}
\centering
\tiny
\includegraphics[width=17cm]{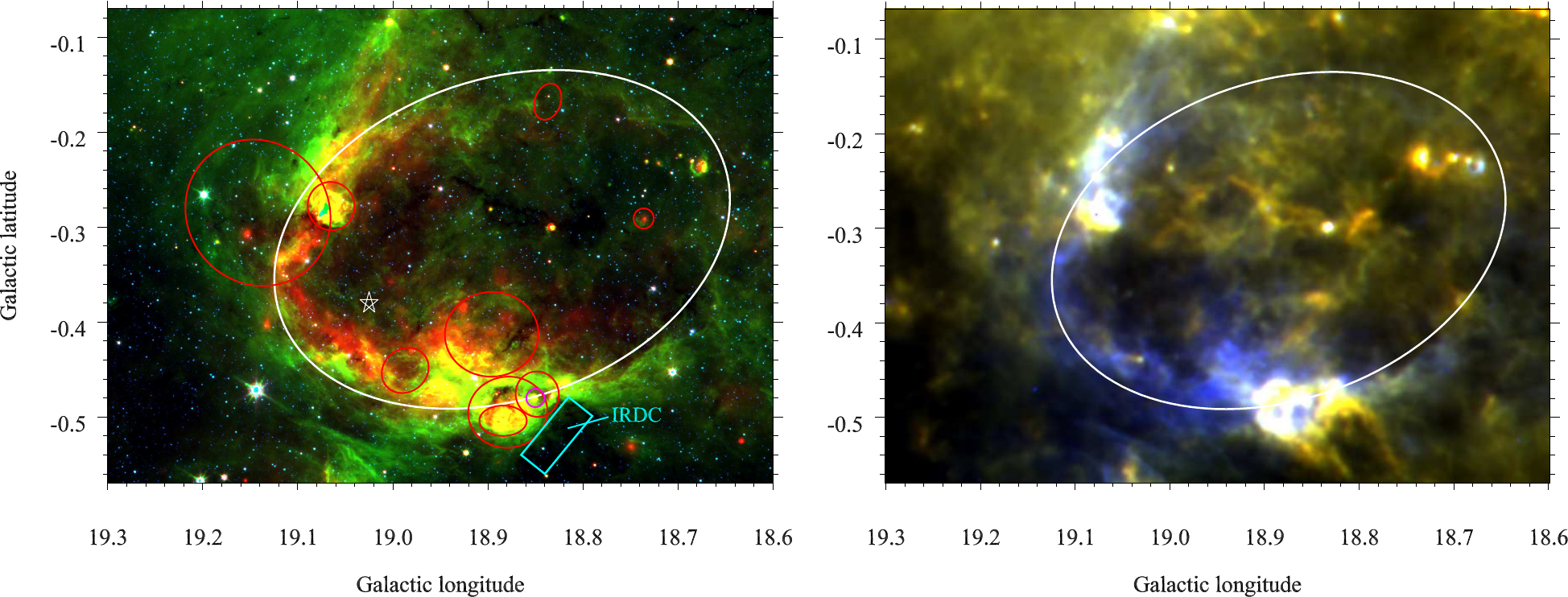}
\vspace{0.5em}
\caption{
{\it Left}: The mid-IR emission of bubble N\,24: composite three-colour image made from \textit{Spitzer} GLIMPSE and
MIPSGAL data (4.5\,$\mu$m = blue, 8\,$\mu$m = green, and 24\,$\mu$m = red).
The red and purple ellipses represent the large and small bubbles located in and around the N\,24 shell.
The white pentagram denotes the position of the exciting star(s), which was located in  $l = 19$\fdg$025$,
and $b = -0$\fdg$38$ and has been identified by \citet{ker13}. The cyan box represents the location of the IRDC on the rim of bubble.
{\textit Right}: The far-IR emission of bubble N\,24: a \textit{Herschel} three-colour image from PACS and SPIRE data
(70\,$\mu$m = blue, 250\,$\mu$m = green, and 350\,$\mu$m = red).
The white ellipse shows the approximate extent of the bubble, which is best seen in infrared images.}
\label{fig1}
\end{figure*}

N\,24 is a large IR bubble in the inner Galaxy at $l = 18$\fdg$908$,
and $b = -0$\fdg$315$ with an effective radius of 10$\arcmin$.93 \citep{sim12},
corresponding to $\sim$ 15 pc at a distance of 4.6 kpc \citep{deh10}.
In a recent detailed study, \citet[W\,39,][]{ker13} compared H\,{\sc{i}} absorption spectra and
Galactic rotation models and found a velocity $V_{\mathrm{LSR}}$ $\sim$ $64.5\pm0.5$
km s$^{-1}$ for this region and confirmed an average (near) kinematic distance of $4.5\pm0.2$ kpc.
These authors also estimated an evolutionary age of ~10$^{6}$ years for this H\,{\sc{ii}} region.
The hierarchical bubble morphology is best seen in the mid-infrared (left image of Fig. \ref{fig1}).
Enclosed by an 8.0\,$\mu$m shell that is generally produced by the absorption of far-ultraviolet photons from the
H\,{\sc{ii}} region by polycyclic aromatic hydrocarbons (PAHs), the extended 24\,$\mu$m emission inside the bubble
is believed to result from hot dust grains that have been heated by absorption of Lyman continuum photons near
the central ionized stars.
Several 8\,$\mu$m absorption features inside and on the rim of the bubble indicate infrared dark cloud (IRDC) structures that have been cataloged in the IRDC catalogs of \citet{sim06}. The catalogued IRDCs inside the bubble appear to have little to do with N\,24, as discussed in Section \ref{3.3.3}. The single IRDC in the distorted section of the rim (the cyan rectangle on the left of Fig. \ref{fig1}) is bordered by bright 8\,$\mu$m features that appear to be the result of the bubble expanding and interacting with it. In the right panel of Figure \ref{fig1}, the IRDCs are clearly visible in the \textit{Herschel} data, which indicates that these structures most likely trace the collected neutral material in cold and dark molecular clouds. The advantage of studying the dust properties of the entire bubble with a widespread wavelength coverage (from 70 to 500\,$\mu$m) and at high resolution is that the spectral energy distribution (SED) of young stellar object candidates (YSOs) at earlier phases can be better constrained and the physical parameters can be more accurately estimated.

In this work, we adopt the numerical results of \citet{ker13} for N\,24 and add far-infrared observations carried out by the \textit{Hershel Space Observatory} and ammonia data
observed from \textit{Nanshan Observatory}. We use these data to further investigate the interaction of the
bubble with its surroundings by studying the dense clumps around it and to explore the possible scenario
of triggered star formation.
The data are described in Section \ref{sec2},
and the results and analysis are presented in Section \ref{sec3} with a summary in Section \ref{sec4}.

\section{Observations and Data Sets}
\label{sec2}
\subsection{Archive data}

Mid-infrared data of the region towards N\,24 have been obtained from the archives of the $\textit{Spitzer Space
Telescope}$ \citep{wer04}.
Images in four IRAC bands centered at 3.6, 4.5, 5.8, and 8.0\,$\mu$m with spatial resolutions ranging from
1$\arcsec$.7 to 2$\arcsec$ have been taken from the Galactic Legacy Infrared Midplane Survey Extraordinaire
(GLIMPSE) survey \citep{ben03}.
The 24\,$\mu$m image with a resolution of 6$\arcsec$ has been obtained from the MIPS Galactic Plane Survey
(MIPSGAL) survey \citep{rie04, car09}.
These data have been used to investigate the structure of the bubble and the classification of the clumps.

Far-infrared data have been obtained from the ESA $\textit{Herschel Space Observatory}$ Infrared Galactic
(Hi-GAL) Plane Survey.
The Hi-GAL plane survey is an Open Time Key Project \citep{pil10, mol10} that mapped the Inner
Galactic plane at 70 and 160\,$\mu$m with the Photoconductor Array Camera and Spectrometer \citep[PACS,][]{pog10}
and at 250, 350, and 500\,$\mu$m with the Spectral and Photometric Imaging
Receiver \citep[SPIRE,][]{gri10}.
The spatial resolution of the images are 6$\arcsec$, 12$\arcsec$, 18$\arcsec$, 24$\arcsec$, and 35$\arcsec$
for the five wavelength bands \citep{mol10}, respectively.
These data are used to investigate the physical properties of cold dust emission associated with N\,24.

Sub-mm data have been obtained from the Atacama Pathfinder Experiment Telescope (APEX) Large Area Survey
of the Galaxy (ATLASGAL) \citep{sch09}.
The ATLASGAL is an unbiased 870\,$\mu$m sub-millimeter survey of the Inner Galactic plane carried out with
the Large APEX Bolometer Camera \citep[LABOCA,][]{sir09}.
At this wavelength, the APEX Telescope has a full width at half-maximum (FWHM) beam size of 19$\arcsec$.2.
Using this data, we investigate the physical properties of the cold dust clumps.
In addition, we used the 20\,cm radio continuum emission data obtained from the Multi-Array Galactic Plane
Imaging Survey (MAGPIS) with an angular resolution of about 5$\arcsec$ \citep{hel06}.

To trace the molecular gas, we used the $^{13}$CO\,(1-0) observations extracted from the Boston University (BU)
and the Five College Radio Astronomy Observatory (FCRAO) Galactic Ring Survey \citep[GRS,][]{jac06}.
The angular resolution is 46$\arcsec$ and the velocity resolution is 0.21\,km s$^{-1}$ covering a range from -5 to +135\,km s$^{-1}$.

\subsection{Molecular Observations}

In order to survey the molecular emission, we observed NH$_{3}$\,(1,1) and (2,2) line emission using the
Nanshan 26 meter radio telescope (NSRT) of the XinJiang Astronomical Observatory, Chinese Academy of
Sciences in March 2018.
The rest frequency was set at 23.708564 GHz for observing NH$_{3}$\,(1,1) at 23.694495\,GHz and
NH$_{3}$\,(2,2) at 23.722633\,GHz, simultaneously.
The typical system temperature is about 50\,K, the FWHM beam of the telescope is 115$\arcsec$ obtained
from point-like continuum calibrators, and the velocity resolution is 0.098\,km s$^{-1}$ provided by an
8192 channel Digital Filter Bank in a 64\,MHz bandwidth mode.
Mapping observations were made using the on-the-fly (OTF) observing mode with a 6$\arcmin$ $\times$ 6$\arcmin$
grid size and a 30$\arcsec$ ($\frac{1}{4}$ beam) sample step.
The antenna temperature corrected for atmospheric attenuation given by the system was $T_{\mathrm{a}}^{*}$,
whose values were calibrated against periodically (6\,s) injected signals from a noise diode.
The main-beam brightness temperature was derived using the relationship
$T_{\mathrm{mb}}=T_{\mathrm{A}}^{*}/\eta_{\mathrm{mb}}$, where the main beam efficiency
$\eta_{\mathrm{mb}}$ is 0.66. The telescope pointing and tracking accuracy is better than 18$\arcsec$.
The $T_{\mathrm{mb}}$ scale has an estimated calibration uncertainty of about 14 \% \citep{wu18}.

\section{Results and Analysis}\label{sec3}
\subsection{Ionized Emission}

\begin{figure}
\centering
\tiny
\includegraphics[width=0.48\textwidth]{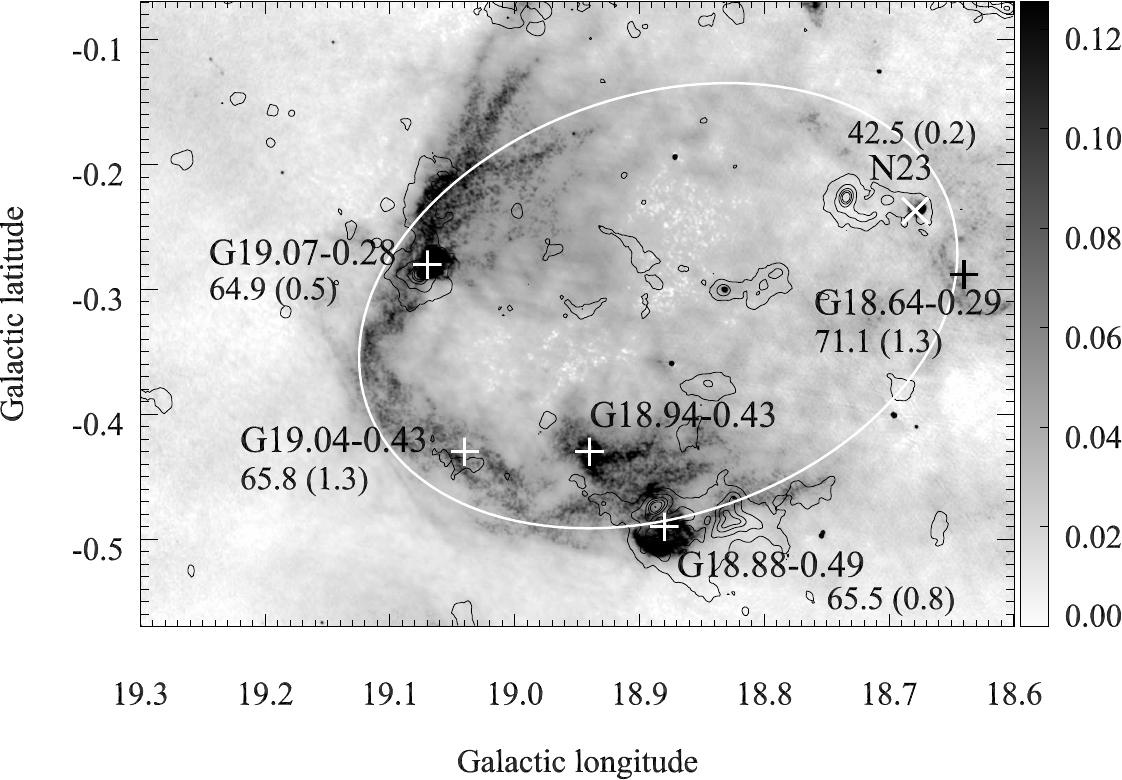}
\caption{The ATLASGAL 870 $\mu$m contours superimposed on the MAGPIS 20\,cm continuum image of the N\,24 region ($\sigma$=0.06\,Jy/beam; contour levels 3, 10, 20, 30,40, 60, 90\,$\sigma$). The grey-scale color bar is in units of Jy/beam. Crosses indicate the positions of RRL observations from \citet{loc89}, together with the peak $V_{\mathrm{LSR}}$ (km s$^{-1}$) with its uncertainty. The position of the H\,{\sc{ii}} region G\,18.94-0.43 given by \citet{ker13} is also shown.
The N\,23 region and its associated clumps clearly have no association with N\,24. The white ellipse is the same as shown in Figure \ref{fig1}.}
\label{fig2}
\end{figure}

Figure \ref{fig2} shows the contour map of the ATLSAGAL 870\,$\mu$m emission superimposed on the MAGPIS
radio-continuum image at 20\,cm, along with the position and velocity of cm-wavelength observations of the RRLs
H85$\alpha$, H87$\alpha$, and H88$\alpha$ \citep{loc89}.
Sub-millimeter waveband 870\,$\mu$m emission is a tracer of cold dust, and 20\,cm emission
is generally considered to be free-free continuum emission from ionized gas, which can be used to trace the
ionising influence of the H\,{\sc{ii}} region.
The ionised gas in N\,24 is spatially associated with the hot dust grains and is
surrounded by a ring of PDRs indicating the strong effects of the H\,{\sc{ii}} region on its surrounding.
Four compact H\,{\sc{ii}} regions (G\,18.88-0.49, G\,18.94-0.43, G\,19.04-0.43, and G\,19.07-0.28) in Figure \ref{fig2} are found to lie on the
ridge of emission.
In addition, more bubbles or H\,{\sc{ii}} regions have been identified on the periphery of N\,24 by the more than 3,500
astronomical volunteers and experts \citep{sim12} in the 8\,$\mu$m $\textit{Spitzer}$-IRAC
and 24\,$\mu$m $\textit{Spitzer}$-MIPSGAL data (see the left of Fig. \ref{fig1}. Red and purple ellipses
represent the large and small bubbles, respectively).
The structure and age of N\,24 are consistent with triggered star formation. The long evolution of N\,24 would have provided enough time for sweeping up the surrounding material and for the mass in the shell to reach the extent, where gravitational instability resulted in the collapse of new stars.
The newly formed stars then ionised their surrounding matter to form new H\,{\sc{ii}} regions.
\citet{ker13} roughly estimated the age of the G\,19.07-0.28 region at about 10$^{5}$ years, and they
concluded that the probable timescale for the expansion of N\,24 and the likely ages of the
secondary regions of massive star formation are consistent with a scenario of triggered star formation.

\subsection{Dust Emission}

\begin{figure}
\centering
\tiny
\includegraphics[width=0.48\textwidth]{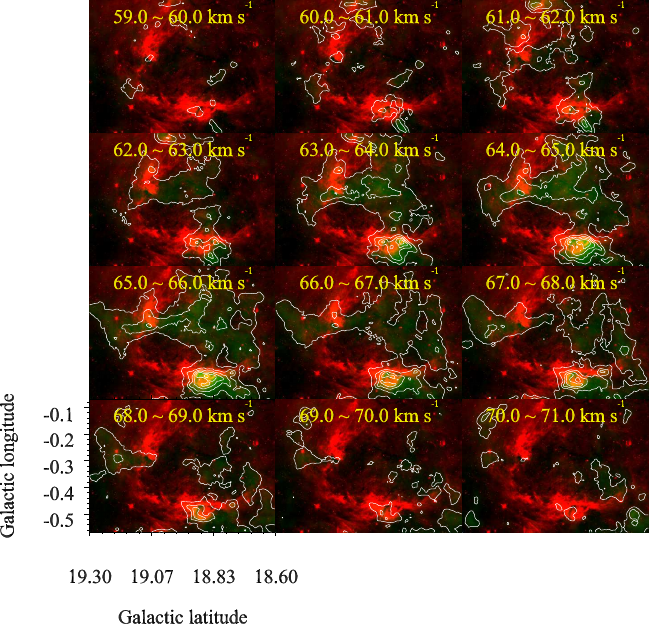}
\caption{Integrated velocity channel maps of the GRS $^{13}$CO\,(1-0) emission at 1.0\,km s$^{-1}$ intervals
from 59.0 to 71.0\,km s$^{-1}$ (in green) superimposed on the 8\,$\mu$m emission (in red).
The contour levels of the $^{13}$CO\,(1-0) emission are 1 to 11.5\,K km s$^{-1}$ by 1.5\,K km s$^{-1}$.}
\label{fig3}
\end{figure}

Because N\,24 is large and contains complex molecular structure, we show the integrated velocity channel
maps of the GRS $^{13}$CO\,(1-0) emission in this region from 59.0 to 71.0\,km s$^{-1}$ in steps of 1.0\,km s$^{-1}$
in order to better determine the corresponding velocities of different components (see Fig. \ref{fig3}).
Two molecular condensations that are spatially associated with small H\,{\sc{ii}} regions are detected at
60.0 to 69.0 km s$^{-1}$ in the East and South of the shell, suggesting that they are physically related to N\,24.
Figure \ref{fig2} shows that the distribution of 870\,$\mu$m emission is consistent with that of molecular emissions.
These condensations are associated with the H\,{\sc{ii}} regions G\,18.88-0.49 and G\,19.07-0.28 and
form a shell surrounding the central H\,{\sc{ii}} region, suggesting that this H\,{\sc{ii}} region is expanding and
sweeping up surrounding material and that feedback from massive stars is taking place.
However, several molecular emissions are also detected at 63.0 to 68.0\,km~s$^{-1}$ inside the bubble, which are
associated with IRDCs and may be not directly related to N\,24 (see Section \ref{3.3.3}).

\subsubsection{Dust Temperature and Column Density Distributions}\label{3.2.1}

Assuming optically thin dust emission, we studied the properties of dust emission using the $\textit{Herschel}$ data
for four bands at 160, 250, 350, and 500\,$\mu$m.
To obtain the distribution of the dust temperature, a pixel-by-pixel spectral energy distribution (SED) fitting has been
 performed on the data.
First, we convolved and regridded the 160, 250, and 350\,$\mu$m images to the lowest resolution 35$\arcsec$ and
largest pixel size 11$\arcsec$.5 of the 500\,$\mu$m image.
Then a grey body function \citep{war90} for a single temperature can be expressed
as follows:
 \begin{equation}\label{eq-1}
          F_{\nu} = \Omega B_{\nu}(T_{\mathrm{d}})(1-e^{-\tau}),
          \end{equation}
where $\Omega$ is the effective solid angle corresponding to each pixel, $B_{\nu}(T_{\mathrm{d}})$ is the
Plank function at the dust temperature $T_{\mathrm{d}}$, which is defined as $B_{\nu}(T_{\mathrm{d}}) =
(2h\nu^{3}/ c^{2})[\mathrm{exp}(h\nu/kT)-1]^{-1}$, where $\textit{h}$ is the Planck constant, $\nu$ is the
frequency, $\textit{c}$ is the speed of light, and $\textit{k}$ is the Boltzmann constant.
The optical depth $\tau$ is given by the relation $\tau = (\nu/ \nu_{c})^{\beta}$, where $\beta = 2$  is the
dust emissivity index, which is a statistical value found in a large sample of H\,{\sc{ii}} regions \citep{and12}, and
$\nu_{c}$ is the critical frequency at which $\tau = 1$.
The two free parameters used in the fitting are the dust temperature $T_{\mathrm{d}}$ and the critical frequency
$\nu_{c}$. We assigned an uncertainty of 20\,$\%$ to each \textit{Herschel} flux and use it as a calibration error \citep{fai12}.
The quality of the SED fitting is assessed via a $\chi^{2}$ minimisation, by considering the observed flux at each
the four Hi-GAL wavebands available for every individual association.

\begin{figure*}
\centering
\tiny
\includegraphics[width=17cm]{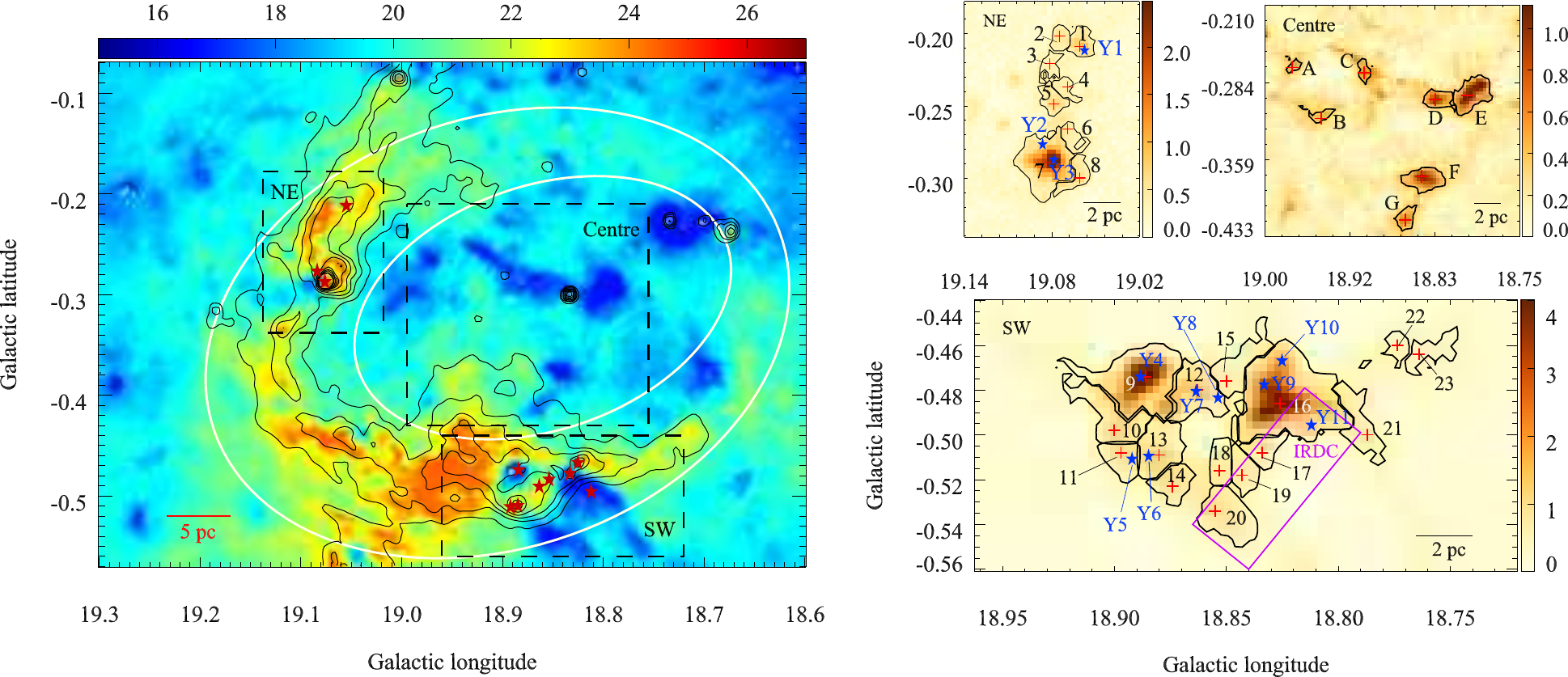}
\caption{ {\textit Left}: The map of the dust temperature determined by pixel by pixel SED fitting.
The black contours define the 70\,$\mu$m emission and several crimson pentagrams represent the positions
of point sources associated with clumps.
The white ellipses approximately trace the profile of the N\,24 shell.
{\textit Right}: Three maps of the H$_{2}$ column density corresponding to the three positions on the left.
Frames NE and SW represent the region on the shell, while frame Centre represents the region of IRDCs inside the bubble.
The identified clumps are shown as thick black contours,
The red crosses represent the densest position of these clumps.
The blue pentagrams denote the point sources shown in the left frame. The purple box represents the location of the IRDC which is the same as shown in Figure \ref{fig1}.
The color bars of left and right figures are given in units of K and 10$^{22}$cm$^{-2}$, respectively.}
\label{fig4}
\end{figure*}

Subsequently the ATLSAGAL 870 $\mu$m image was convolved and regridded to the lowest resolution and
largest pixel size of the 500 $\mu$m image.
The hydrogen molecular column density distribution of N\,24 was then calculated pixel by pixel:
              \begin{equation}\label{eq-2}
               N_{\mathrm{H}_{2}}={{S_{\nu}}R\over{B_\nu}(T_{\mathrm{d}}){\Omega}{\kappa_{\nu}}{\mu}{m_{\mathrm{H}}}},
              \end{equation}
where $S_{\nu}$ is 870\,$\mu$m flux density and $m_{\mathrm{H}}$ is the mass of an hydrogen atom.
$\mu$ is the mean molecular weight of the interstellar medium, $\textit{R}$ is the dust-to-gas ratio, $\kappa_{\nu}$
is the dust absorption coefficient, which we assume to be equal to 2.8, 100, 1.85\,cm$^{2}$g$^{-1}$, respectively
\citep[i.e.,][]{kau08, sch09, cse14}.
The resulting map of the dust temperature in Figure \ref{fig4} shows dust heating from the interior to the edge
of N\,24, and nearly matches with the contours for the 70\,$\mu$m emission.
The shell morphology seen at mid-IR wavebands and the relatively higher dust temperature in the shell of the
bubble suggest that the large H\,{\sc{ii}} region N\,24 has been heating its surroundings.
In spite of the second-generation bubbles distributed on the shell mixed with the large H\,{\sc{ii}} region making
the temperature distribution appear complex, the higher temperatures are found mostly in the
diffuse regions heated by the H\,{\sc{ii}} regions and areas associated with YSOs in the shell.
The anti-correlation between $N_{\mathrm{H_{2}}}$ and $T_{\mathrm{dust}}$ may be attributed to the dense cores
being formed by cold dust where the external heating from the H\,{\sc{ii}} region is lower.

\subsubsection{Dense Clumps Identification and Physical Parameters}\label{3.2.2}

Based on the distribution of hydrogen molecular column density, the Clumpfind2d algorithm was used to identify
dense clumps \citep{wil94}.
First, the root mean square (rms) of the column density map was derived based on a background noise estimate
of 4 $\times$ 10$^{20}$ cm$^{-2}$ from weak-emission regions.
Then the level range from a threshold of 4\,$\delta$ to 20\,$\delta$ was set with increments of 2.5\,$\delta$ (where,
$\delta$ is equal to the rms noise level), which is similar to the previous parameters used \citep[i.e.,][]{lu14, oha16}.
Twenty-three possible dense clumps associated with N\,24 and seven possible dense clumps associated with IRDCs inside the bubble were identified within the $N_{\mathrm{H_{2}}}$ map. These results match to the greatest extent the dense clumps identified by \citet{urq18} using the GaussClump algorithm within the ATLASGAL 870\,$\mu$m emission map. Twenty-three of our 30 clumps have a counterpart while clumps 3, 4, 18, 19, 22, 23, B have no counterparts. Eighteen clumps are within an angular distance of 15 arcsec, of which twelve are in the shell and six are associated with IRDCs.
\citet{pin09} queried different parameter settings in the Clumpfind2d algorithm for slightly different results and it works consistently for relatively isolated clumps.
Adopting the output from the Clumpfind2d algorithm it should be noted that the clumps identified
around N\,24 are located in three distinct regions NE, Centre, and SW in Figure \ref{fig4} (left).
The three pictures in Figure \ref{fig4} (right) show the relative position and size of the identified clumps.
The two regions NE and SW in the bubble shell are also the main regions where \citet{ker13} studied
YSOs, suggesting that the expanding H\,{\sc{ii}} region formed dense molecular clouds and further triggered
star formation.
It should be noted that the clumps 16, 17, 19, 20, and 21 are coincident with the IRDC at the SW edge of the bubble in Figure \ref{fig4} (right). This IRDC probably existed before bubble expanded, so we cannot ignore the possibility that these clumps and YSOs are part of the same star forming environment that formed N\,24, and the clumps would have formed stars without the influence of the bubble.
The seven clumps in the Centre region are associated with IRDCs and appear to have little to do with the
expanding H\,{\sc{ii}} region and will not be discussed further in this paper.

In order to determine the evolutionary stages of the 23 clumps in the NE and SW regions, a visual inspection was done using the mid-infrared images (3.6, 4.5, 8.0, and 24\,$\mu$m), and the MAGPIS 20 cm \citep{hel06}, Hi-GAL 70\,$\mu$m \citep{pog10} emission images as described by \citet{fos11} and \citet{hoq13}. To increase the reliability of this determination, the 24 $\mu$m highly reliable point sources from the MIPSGAL catalog were also used to match the clumps \citep[i.e.,][]{he16}.
All 23 clumps were assigned a consecutive evolutionary stage: quiescent (pre-stellar), proto-stellar, and
H\,{\sc{ii}} region.
In principle, a pre-stellar clump appears dark at 3.6, 4.5, 8.0, 24, and 70\,$\mu$m, and has no embedded massive
YSOs.
Proto-stellar clumps have 24\,$\mu$m point sources, 70\,$\mu$m compact emission or
are associated with extended 4.5\,$\mu$m emission that is likely associated with shocks generated by molecular
outflows \citep{hoq13}.
Clumps that have 20\,cm emission and are associated with a compact H\,{\sc{ii}} region that emits
extended 8.0, and 24\,$\mu$m emissions are also classified as H\,{\sc{ii}}.
However, in practise the size of the H\,{\sc{ii}} regions around N\,24 are so large that they can not be
designated as single clumps. Therefore clumps that are close to the center of a H\,{\sc{ii}} region,
but do not contain the whole H\,{\sc{ii}} region and do not have a 24\,$\mu$m point source, will still be categorised as an H\,{\sc{ii}} region.
The classification is shown in column 10 of Table \ref{t1}.

\begin{figure}
\centering
\tiny
\includegraphics[width=0.48\textwidth]{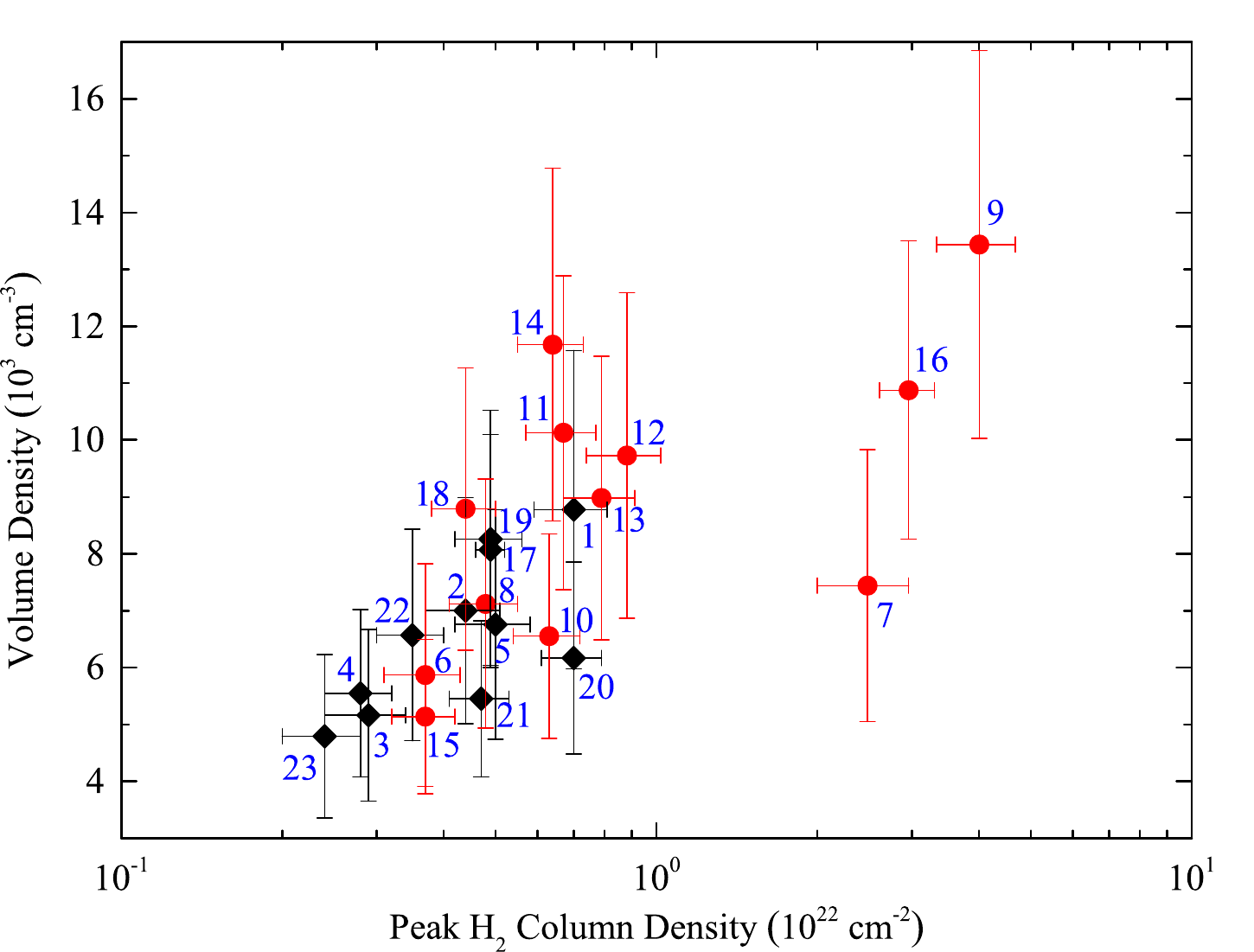}
\caption{The volume density versus the peak H$_{2}$ column density for 23 identified clumps.
The red circles represent the clumps around second-generation bubbles, while the black diamonds represent
clumps surrounding the large H\,{\sc{ii}} region of N\,24.}
\label{fig5}
\end{figure}

Assuming optically thin dust emission, the clump masses were derived using a similar method as used for
determining the H$_{2}$ column density via the formula:
            \begin{equation}\label{eq-3}
             M={{D^2}{S_{\nu}}{R}\over{B_\nu}(T_{\mathrm{d}}){\kappa_\nu}},
            \end{equation}
where $\textit{D}$ is the heliocentric distance to the clumps, $S_{\nu}$ is the 870\,$\mu$m flux density, $B_{\nu}
(T_{\mathrm{d}})$ is the Plank function for a dust temperature $T_{\mathrm{d}}$, and the parameter settings
of $\textit{R}$ and $\kappa_{\nu}$ are the same as those in Section \ref{3.2.1}.
Assuming that each clump has a spherical geometry, the average volume density may be derived as:
               \begin{equation}\label{eq-4}
               n_{\mathrm{H}_{2}}=3M/(4{\pi}{R_{\mathrm{eff}}^{3}}{\mu}{m_{\mathrm{H}}})
                \end{equation}
where $\textit{M}$ is the mass that was derived from the 870\,$\mu$m flux, $m_{\mathrm{H}}$ is the mass of
an hydrogen atom, $R_{\mathrm{eff}}$ is the effective radius of the structure output from the Clumpfind2d algorithm, and the mean molecular weight $\mu$ is
assumed to be 2.8.
Finally, a summary of the aforementioned physical parameters of 23 clumps are given in Cols. 5-9 of Table \ref{t1}.
The effective radius, peak dust temperature, peak column density, mass, and volume density of these molecular
clumps fall in the range of 0.65-1.73\,pc, 16.3-24.2\,K, 0.24-4.01 $\times$ 10$^{22}$ cm$^{-2}$,
0.60-16.30 $\times$ 10$^{3}$ M$_{\odot}$, 4.79-13.44 $\times$ 10$^{3}$ cm$^{-3}$,
respectively.
And the mean values of these parameters are ~0.92 $\pm$ 0.06 pc, ~20.8 $\pm$ 0.5 K, ~0.86 ( $\pm$ 0.19) $\times$ 10$^{22}$ cm$^{-2}$, ~2.66 ( $\pm$ 0.81) $\times$ 10$^{3}$
M$_{\odot}$, ~7.75 ( $\pm$ 0.46) $\times$ 10$^{3}$ cm$^{-3}$, respectively.
Figure \ref{fig5} shows the distribution of the volume density versus the peak H$_{2}$ column density for the 23
clumps.
Compared with the clumps around the edge of the large bubble, the clumps around second-generation bubbles tend to
have a larger column density and volume density, which suggests that these bubbles provided additional feedback
making the clumps more dense.

  \begin{table*}

  \caption{Characteristics of twenty-three dust clumps.}
  \label{t1}
  \begin{tabular}{cccccccccc}
    \hline\hline
    Number & Clump name & R.A. & Decl. & $R_{\mathrm{eff}}$ & $T_{\mathrm{dust, peak}}$ & $N_{\mathrm{H_{2}, peak}}$ & $M_{\mathrm{clump}}$ &$ n_{\mathrm{H_2}}$ & Stage  \\
        &    & J2000 (deg) & J2000 (deg) & (pc) & (K) & (10$^{22}$cm$^{-2}$) & 10$^{3}$M$_{\odot}$ &
        10$^{3}$cm$^{-3}$ &       \\
    \hline
    1   & G\,19.058-00.209 &	276.62293&	-12.419557&	0.81$\pm$0.04&	23.1$\pm$2.6&	0.70$\pm$0.11&	1.35$\pm$0.23&	8.78$\pm$2.80&	Proto-stellar    \\
    2   & G\,19.072-00.202 &	276.62327&	-12.403908&	0.72$\pm$0.03&	22.5$\pm$2.5&	0.44$\pm$0.07&	0.76$\pm$0.12&	7.00$\pm$1.99&	Pre-stellar	     \\
    3   & G\,19.079-00.221 &	276.64382&	-12.406570&	0.77$\pm$0.03&	24.2$\pm$3.0&	0.29$\pm$0.05&	0.68$\pm$0.12&	5.16$\pm$1.51&	Pre-stellar	     \\
    4   & G\,19.067-00.237 &	276.65235&	-12.425086&	0.75$\pm$0.03&	21.7$\pm$2.4&	0.28$\pm$0.04&	0.68$\pm$0.11&	5.55$\pm$1.47&	Pre-stellar	     \\
    5	& G\,19.076-00.249 &	276.66776&	-12.422273&	0.85$\pm$0.04&	21.8$\pm$2.4&	0.50$\pm$0.08&	1.20$\pm$0.19&	6.76$\pm$2.02&	Pre-stellar	     \\
    6	& G\,19.067-00.266 &	276.67887&	-12.438157&	0.80$\pm$0.04&	24.0$\pm$3.0&	0.37$\pm$0.06&	0.87$\pm$0.16&	5.87$\pm$1.96&	H\,{\sc{ii}}	 \\
    7	& G\,19.079-00.289 &	276.70543&	-12.438254&	1.64$\pm$0.07&	23.1$\pm$2.9&	2.48$\pm$0.48&	9.49$\pm$1.84&	7.44$\pm$2.39&	H\,{\sc{ii}}     \\
    8	& G\,19.057-00.300 &	276.70490&	-12.462845&	0.84$\pm$0.04&	22.5$\pm$2.5&	0.48$\pm$0.07&	1.22$\pm$0.20&	7.12$\pm$2.19&	Pre-stellar	     \\
    9	& G\,18.888-00.476 &	276.78006&	-12.695805&	1.41$\pm$0.06&	17.0$\pm$1.6&	4.01$\pm$0.67&	10.87$\pm$1.37&	13.44$\pm$3.41&	Proto-stellar    \\
    10	& G\,18.900-00.498 &	276.80953&	-12.693981&	1.07$\pm$0.05&	20.4$\pm$2.0&	0.63$\pm$0.09&	2.32$\pm$0.31&	6.55$\pm$1.80&	H\,{\sc{ii}}	 \\
    11	& G\,18.897-00.508 &	276.81717&	-12.701291&	0.73$\pm$0.03&	21.2$\pm$2.2&	0.67$\pm$0.10&	1.14$\pm$0.17&	10.13$\pm$2.76&	H\,{\sc{ii}}	 \\
    12	& G\,18.863-00.480 &	276.77555&	-12.718347&	0.99$\pm$0.04&	22.7$\pm$2.5&	0.88$\pm$0.14&	2.73$\pm$0.47&	9.73$\pm$2.86&	Proto-stellar    \\
    13	& G\,18.880-00.509 &	276.80997&	-12.716802&	0.93$\pm$0.04&	21.6$\pm$2.3&	0.79$\pm$0.12&	2.09$\pm$0.31&	8.98$\pm$2.49&	H\,{\sc{ii}}	 \\
    14	& G\,18.875-00.521 &	276.81847&	-12.726813&	0.69$\pm$0.03&	19.5$\pm$1.8&	0.64$\pm$0.09&	1.11$\pm$0.15&	11.68$\pm$3.10&	Pre-stellar	     \\
    15	& G\,18.849-00.474 &	276.76342&	-12.727943&	0.98$\pm$0.04&	20.2$\pm$2.0&	0.37$\pm$0.05&	1.40$\pm$0.20&	5.14$\pm$1.36&	Pre-stellar	     \\
    16	& G\,18.824-00.486 &	276.76237&	-12.755654&	1.73$\pm$0.08&	16.3$\pm$1.2&	2.96$\pm$0.35&	16.30$\pm$1.67&	10.88$\pm$2.62&	Proto-stellar    \\
    17	& G\,18.836-00.508 &	276.78391&	-12.756535&	0.72$\pm$0.03&	18.7$\pm$1.6&	0.49$\pm$0.03&	0.87$\pm$0.11&	8.07$\pm$2.03&	Pre-stellar	     \\
    18	& G\,18.852-00.515 &	276.80205&	-12.744376&	0.65$\pm$0.03&	19.9$\pm$1.9&	0.44$\pm$0.06&	0.70$\pm$0.10&	8.79$\pm$2.48&	Pre-stellar      \\
    19	& G\,18.842-00.515 &	276.79728&	-12.753226&	0.67$\pm$0.03&	19.3$\pm$1.8&	0.49$\pm$0.07&	0.72$\pm$0.10&	8.26$\pm$2.26&	Pre-stellar      \\
    20	& G\,18.856-00.531 &	276.81848&	-12.748283&	1.03$\pm$0.05&	17.7$\pm$1.5&	0.70$\pm$0.09&	1.95$\pm$0.25&	6.17$\pm$1.69&	Pre-stellar      \\
    21	& G\,18.788-00.499 &	276.75698&	-12.793565&	0.94$\pm$0.04&	17.6$\pm$1.5&	0.47$\pm$0.06&	1.31$\pm$0.16&	5.45$\pm$1.37&	Pre-stellar      \\
    22	& G\,18.776-00.460 &	276.71586&	-12.786022&	0.68$\pm$0.03&	21.0$\pm$2.2&	0.35$\pm$0.05&	0.60$\pm$0.09&	6.57$\pm$1.86&	Pre-stellar      \\
    23	& G\,18.763-00.467 &	276.71601&	-12.800786&	0.84$\pm$0.04&	21.6$\pm$2.3&	0.24$\pm$0.04&	0.82$\pm$0.13&	4.79$\pm$1.44&	Pre-stellar      \\
    \hline
  \end{tabular}

 \end{table*}

\subsection{Molecular Line Emission}
\subsubsection{Kinetic temperature}\label{3.3.1}

\begin{figure}
\centering
\tiny
\includegraphics[width=0.48\textwidth]{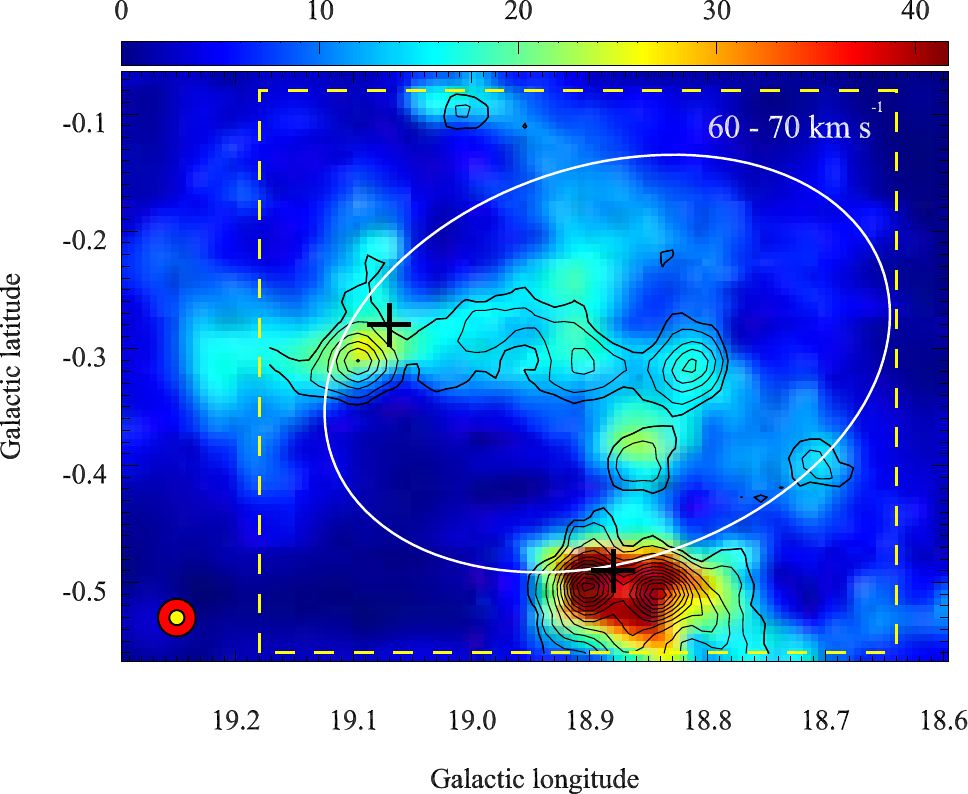}
\caption{Velocity-integrated intensities of the NH$_{3}$\,(1,1) emission overlaid on the integrated velocity
map of the GRS $^{13}$CO\,(1-0) emission, both integrated over 60 to 70\,km~s$^{-1}$.
The contour levels of the ammonia emission start at 5$\sigma$ with steps of 4$\sigma$ ($\sigma$ =
0.05\,K km s$^{-1}$).
The color bar is given in units of K km s$^{-1}$. The observation limits of the ammonia are indicated with yellow dashed lines.
The yellow and red filled circles at the bottom left illustrate the half-power beam size of ammonia and
$^{13}$CO, respectively.
The white ellipse designates the main bubble N\,24 as shown in Figure \ref{fig1}.}
\label{fig6}
\end{figure}

Ammonia (NH$_{3}$) inversion transitions are widely used as a spectroscopic tool to study
the physical properties of dense gas in the ISM because of the lack of depletion for the molecule, its favourable
hyperfine structure, and its sensitivity to kinetic temperature \citep{ho83, wal83, taf04}.
The energy levels of NH$_{3}$ K=1 and K=2 are forbidden transitions, so that the intensity ratio of
NH$_{3}$\,(1,1) and (2,2) lines is only sensitive to collisions, thus making the NH$_{3}$ molecule a temperature
tracer. The NH$_{3}$\,(1,1) and (2,2) lines were observed simultaneously.
Figure \ref{fig6} shows the velocity-integrated intensity map of the NH$_{3}$\,(1,1) emission as contours
superimposed on the GRS\,$^{13}$CO\,(1-0) line emission map integrated from 60.0 to 70.0\,km s$^{-1}$.
The integration range of the NH$_{3}$\,(1,1) emission is also from 60.0 to 70.0\,km s$^{-1}$ to cover the
main group of hyperfine components \citep[$\Delta$ F=0,][]{ho83}, considering that the satellite lines
($\Delta$ F=$\pm$1) are usually not accessible and the signal-to-noise (S/N) ratios may reduce if the
velocity-integrated intensity includes the satellite lines.
Figure \ref{fig6} shows that the distribution of ammonia molecules is spatially well correlated with
the distribution of $^{13}$CO emission, while ammonia traces the denser regions.
The emission region located south of the N\,24 bubble with many small bubbles and strong 8\,$\mu$m emission is more prominent than other regions, which indicates that intense star formation activities are ongoing in this region.

Assuming that the emissions from different NH$_3$ energy levels have an equal beam-filling factor, and that the
excitation temperature is constant
along the line-of-sight, and that the radiation processes occur under conditions of local thermodynamic equilibrium
(LTE), then the rotational temperature between the (1,1) and (2,2) inversion transitions can be derived from the
following equation \citep{ho83}:
     \begin{equation}\label{eq-5}
      T_{\mathrm{rot}}(1,2)= \frac{-41.5}{\ln(\frac{-0.282}{\tau_{\rm{m}}(1,1)}\ln (1-\frac{\Delta
      T_{\mathrm{a}}^{*}(2,2)}{\Delta T_{\mathrm{a}}^{*}(1,1)}(1-\mathrm{exp(-\tau_{m}(1,1))}))}
     \end{equation}
where $\Delta T_{\mathrm{a}}^{*}$ of NH$_{3}$\,(1,1) and (2,2) are the observed antenna brightness temperatures
derived using "GAUSS" fitting in GILDAS\footnote{GILDAS (the Grenoble Image and Line Data Analysis System) is an ensemble of routines developed by IRAM to process single antenna observation spectral line data and interference data. \url{http://www.iram.fr/IRAMFR/GILDAS/}}, and $\tau_{\mathrm{m}}(1,1)$ is the main group opacity
derived using the GILDAS "NH$_{3}$(1,1)" fitting method.
Primary polynomial fitting is used for baseline removal in the fitting process.
Subsequently, the kinetic temperature is obtained from the rotational temperature as derived from a fit of
kinetic and rotational temperatures using different Monte Carlo models \citep{taf04}:
      \begin{equation}\label{eq-6}
       T_{\mbox{\tiny kin}}=\frac{T_{\mbox{\tiny rot}}(1,2)}{1-\frac{T_{\mbox{\tiny rot}}(1,2)}{42 {\mathrm K}}\ln\left(1+1.1\,
       {\mathrm exp}\left(\frac{-16 {\mathrm K}}{T_{\mbox{\tiny rot}}(1,2)}\right)\right)},
       \end{equation}
Finally, we used Gaussian fittings to obtain the central velocity, and the velocity width as listed in Cols. 2-6 of
Table \ref{t2}.
The kinetic temperature of these molecular clumps ranges from 19.0 to 29.5\,K with a mean value of 23.5$\pm$0.6 K.
Those values are slightly larger than the dust temperature, indicating that the ammonia traces the gas associated with the cold dust.

\begin{table*}
  \caption{Parameters of the twenty-three dust clumps.}
  \label{t2}
  \begin{tabular}{cccccccccc}
    \hline\hline
    Number & Central velocity & Velocity width & $\tau$ & $T_{\mathrm{rot}}$ & $T_{\mathrm{kin}}$ & $M_{\mathrm{vir}}$ & $M_{\mathrm{clump}}/M_{\mathrm{vir}}$\\
      & (km s$^{-1}$) & (km s$^{-1})$ &   & (K) & (K) & (10$^{3}$M$_{\odot}$) &   \\
    \hline
    1&	62.9$\pm$0.1&	2.5$\pm$0.3&	1.20$\pm$0.56&	18.4$\pm$1.3&	22.1$\pm$2.4&	0.84$\pm$0.24&	1.61$\pm$0.54\\
    2&	62.9$\pm$0.2&   2.7$\pm$0.3&	1.02$\pm$0.65&	22.6$\pm$1.7&	29.5$\pm$3.7&	0.88$\pm$0.23&	0.86$\pm$0.26\\
    3&	65.3$\pm$0.2&	3.4$\pm$0.4&	0.79$\pm$0.55&	20.4$\pm$1.4&	25.5$\pm$2.8&	1.52$\pm$0.41&	0.45$\pm$0.15\\
    4&	65.4$\pm$0.2&	2.7$\pm$0.4&	1.56$\pm$0.64&	18.8$\pm$1.5&	22.8$\pm$2.8&	0.93$\pm$0.31&	0.73$\pm$0.27\\
    5&	65.5$\pm$0.2&	2.8$\pm$0.4&	1.51$\pm$0.63&	21.2$\pm$1.6&	26.9$\pm$3.3&	1.11$\pm$0.36&	1.08$\pm$0.39\\
    6&	65.4$\pm$0.1&	2.6$\pm$0.2&	1.29$\pm$0.37&	21.0$\pm$1.0&	26.5$\pm$2.0&	0.92$\pm$0.19&	0.95$\pm$0.26\\
    7&  65.6$\pm$0.1&	2.8$\pm$0.2&	1.85$\pm$0.19&	17.2$\pm$0.4&	20.2$\pm$0.7&	2.17$\pm$0.39&	4.37$\pm$1.16\\
    8&	64.8$\pm$0.1&	3.1$\pm$0.2&	0.87$\pm$0.35&	18.1$\pm$0.8&	21.6$\pm$1.5&	1.34$\pm$0.23&	0.91$\pm$0.22\\
    9&	65.9$\pm$0.1&	2.6$\pm$0.1&	1.82$\pm$0.16&	20.4$\pm$0.4&	25.3$\pm$0.8&	1.61$\pm$0.19&	6.75$\pm$1.17\\
    10&	66.3$\pm$0.1&	2.4$\pm$0.2&	0.70$\pm$0.31&	21.1$\pm$0.8&	26.6$\pm$1.6&	1.09$\pm$0.22&	2.13$\pm$0.52\\
    11&	66.4$\pm$0.1&	2.6$\pm$0.2&	1.17$\pm$0.38&	20.8$\pm$1.0&	26.1$\pm$2.0&	0.85$\pm$0.16&	1.34$\pm$0.32\\
    12&	65.5$\pm$0.1&	2.5$\pm$0.1&	1.35$\pm$0.23&	18.9$\pm$0.5&	22.8$\pm$0.9&	1.07$\pm$0.12&	2.55$\pm$0.52\\
    13&	66.2$\pm$0.1&	2.4$\pm$0.1&	1.07$\pm$0.22&	17.9$\pm$0.5&	21.3$\pm$0.9&	0.94$\pm$0.11&	2.22$\pm$0.42\\
    14&	65.8$\pm$0.1&	2.6$\pm$0.2&	0.18$\pm$0.37&	19.8$\pm$1.0&	24.4$\pm$1.9&	0.77$\pm$0.15&	1.44$\pm$0.34\\
    15&	65.3$\pm$0.1&	2.3$\pm$0.2&	1.53$\pm$0.27&	16.9$\pm$0.6&	19.6$\pm$1.1&	0.91$\pm$0.19&	1.54$\pm$0.39\\
    16&	65.1$\pm$0.1&	2.6$\pm$0.1&	2.04$\pm$0.11&	17.3$\pm$0.2&	20.3$\pm$0.4&	2.02$\pm$0.24&	8.07$\pm$1.27\\
    17&	65.5$\pm$0.1&	2.4$\pm$0.1&	1.49$\pm$0.21&	16.6$\pm$0.4&	19.3$\pm$0.7&	0.71$\pm$0.09&	1.23$\pm$0.22\\
    18&	65.5$\pm$0.1&	3.9$\pm$0.4&	0.60$\pm$0.27&	18.2$\pm$0.7&	21.8$\pm$1.3&	1.62$\pm$0.40&	0.43$\pm$0.12\\
    19&	65.6$\pm$0.1&	3.7$\pm$0.5&	0.52$\pm$0.26&	20.2$\pm$0.7&	25.1$\pm$1.4&	1.51$\pm$0.47&	0.48$\pm$0.16\\
    20&	65.3$\pm$0.2&	4.9$\pm$0.3&	0.72$\pm$0.39&	17.1$\pm$0.9&	20.1$\pm$1.6&	4.11$\pm$0.70&	0.47$\pm$0.10\\
    21&	65.2$\pm$0.1&	2.8$\pm$0.2&	1.28$\pm$0.26&	16.4$\pm$0.5&	19.0$\pm$0.9&	1.27$\pm$0.23&	1.03$\pm$0.22\\
    22&	66.3$\pm$0.3&	3.3$\pm$0.5&	1.42$\pm$0.65&	20.2$\pm$1.6&	25.1$\pm$3.2&	1.26$\pm$0.43&	0.48$\pm$0.18\\
    23&	66.4$\pm$0.3&	3.7$\pm$0.6&	1.07$\pm$0.87&	22.6$\pm$2.2&	29.4$\pm$4.8&	1.90$\pm$0.70&	0.43$\pm$0.17\\
    \hline
  \end{tabular}
 \end{table*}

 \begin{figure}
\centering
\tiny
\includegraphics[width=0.48\textwidth]{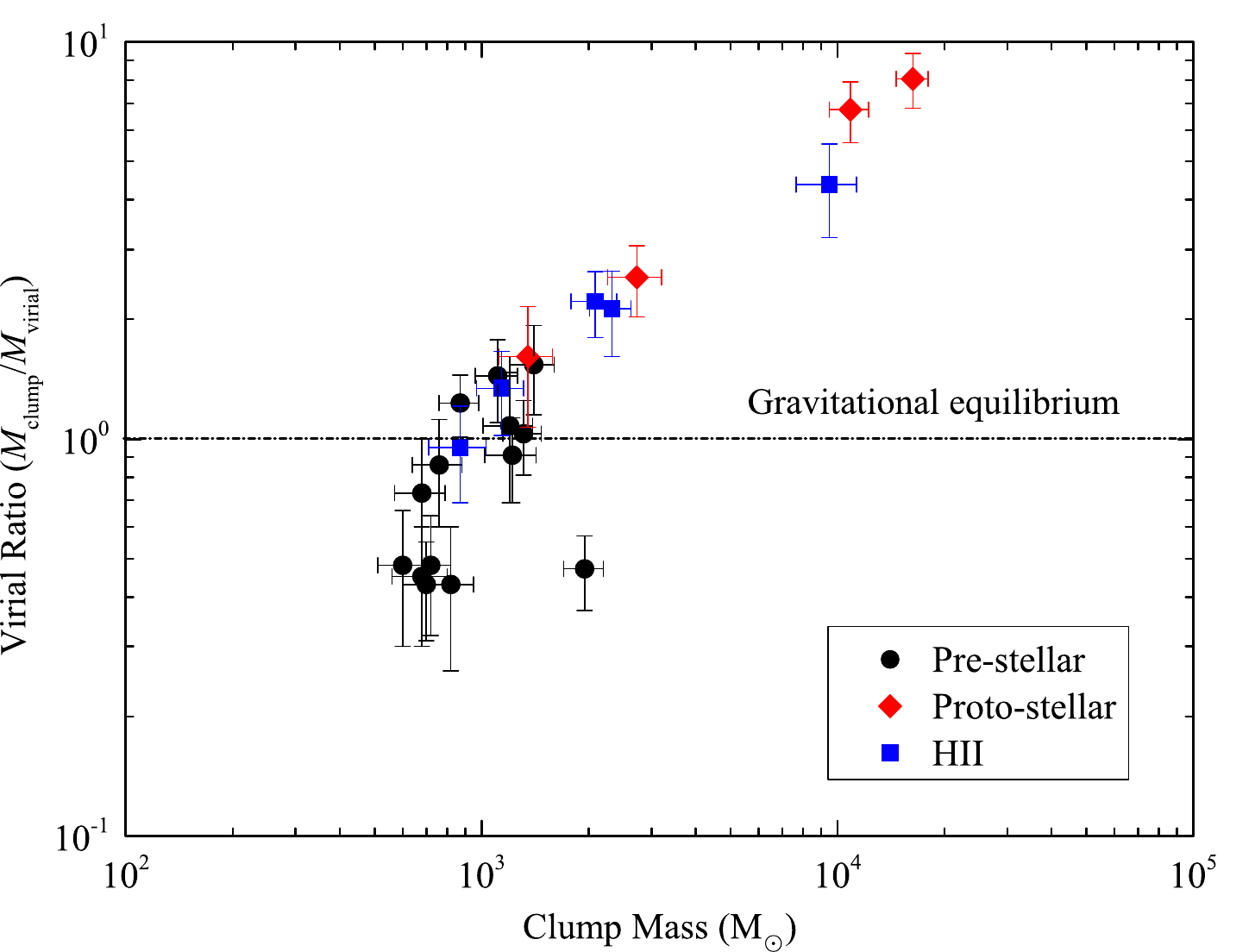}
\caption{Virial Ratio versus the clump mass separated by evolutionary stage.
The black horizontal line denotes the locus of the gravitational equilibrium with thermal and kinematic energy.}
\label{fig7}
\end{figure}

\begin{figure}
\centering
\tiny
\includegraphics[width=0.48\textwidth]{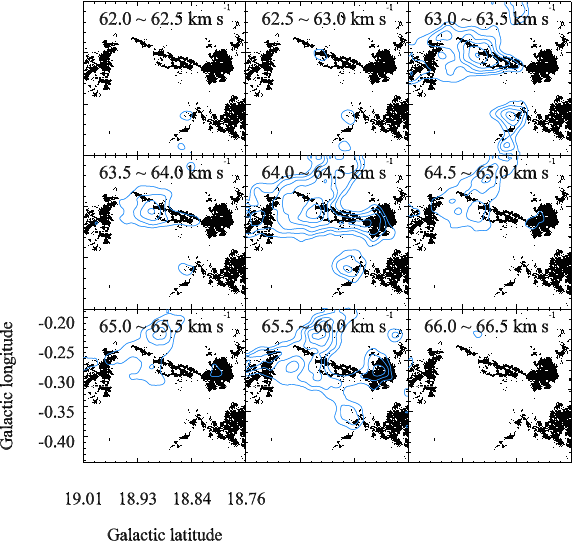}
\caption{Integrated velocity channel maps of the GRS $^{13}$CO\,(1-0) emission every 0.5\,km s$^{-1}$ from
62.0 to 66.5\,km s$^{-1}$ superimposed on the IRDCs that were outlined in the 8\,$\mu$m emission map by setting a threshold below 85\,MJy/sr.
The contour levels of the $^{13}$CO\,(1-0) emission are 1.4 to 2.6\,K km s$^{-1}$ by 0.2\,K km s$^{-1}$.}
\label{fig8}
\end{figure}

\subsubsection{Gravitational stability of clumps}

To assess the gravitational stability of these 23 clumps, we compare the thermal mass and virial mass of the clumps
using the  NH$_{3}$(1,1) emissions.
Assuming a density profile of $\rho \propto r^{-1.8}$ for these resolved clumps, the virial mass can be obtained from the average velocity dispersion \citep{urq13}:
    \begin{equation}\label{eq-7}
    M_{\mathrm{vir}} = \frac{5}{8\ln{2}} \frac{1}{1.3G} R_{\mathrm{eff}} \Delta v_{\mathrm{avg}}^{2},
    \end{equation}
where $R_{\mathrm{eff}}$ is the effective radius of the clump (listed in Table \ref{t1}).
The average velocity dispersion of the total column of gas was estimated from the measured ammonia
line width \citep{ful92} as:
   \begin{equation}\label{eq-8}
    \Delta v_{\mathrm{avg}}^2 = \Delta v_{\mathrm{corr}}^2+8\ln2 \times \frac{\textit{kT}_{\mathrm{kin}}}{\textit{m}_{\mathrm{H}}}(\frac{1}{\mu_{\mathrm{P}}}-\frac{1}{\mu_{\mathrm{NH_{3}}}}),
   \end{equation}
where $\Delta v_{\mathrm{corr}}$ is the observed NH$_{3}$ line width for the resolution of the spectrometer
($\Delta v_{\mathrm{corr}}^{2}=\Delta v_{\mathrm{obs}}^{2}-\Delta v_{\mathrm{channel width}}^{2}$), {\textit k} is
the Boltzmann constant, $\textit{T}_{\mathrm{kin}}$ is the kinetic temperature of the gas, and $\mu_{\mathrm{p}}=2.33$
\citep[i.e.][]{ful92} and $\mu_{\mathrm{NH_{3}}}=17$ are the mean molecular masses of molecular
hydrogen and ammonia, respectively.
The uncertainty in the virial mass comes from errors of the measured distance and source size, as well as the
fitted line width, the spectrometer channel width, and the kinetic temperature.
It should be noted that it has been assumed that the observed NH$_{3}$ line width is representative of the whole
clump when calculating the virial mass, but in reality the line width may decrease towards the edges of the clumps,
which leads to an overestimate \citep{zin97}.
In addition, Equation (\ref{eq-8}) only considers the simplest case of a virial equilibrium where gravity and velocity
width are taken into account, and neglects other influencing factors such as an external pressure and magnetic
field strength.

The estimated virial mass and virial ratio for each clump are presented in Cols. 7-8 of Table \ref{t2} and the
virial ratio is plotted versus the clump mass for our sample of clumps at different stages of evolution in Figure \ref{fig7}.
The virial ratio describes the competition between the internal supporting energy and the gravitational energy.
The dashed black horizontal line in the plot denotes the line of gravitational stability: clumps below this line
are likely to be unbound while those above this line are bound and unstable against gravity.
The mean value of 1.81 and the median value of 1.08 for the virial ratio are both (slightly) larger than unity.
The values for the proto-stellar clumps and almost all H\,{\sc{ii}} clumps are significantly larger than unity, which indicates
that these clumps are gravitationally bound systems and are collapsing to form stars.
On the other hand, only five of 14 pre-stellar sources currently appear to be gravitationally bound systems.
These results show a consistency between our (first order) estimate of the virial ratio and the evolutionary stage of the clumps.

\subsubsection{IRDCs within bubble in line of sight}\label{3.3.3}

Further study on the Centre region of Figure \ref{fig4} using the integrated velocity channel maps of the
$^{13}$CO\,(1-0) emission reveals which of these molecular emissions within the N\,24 bubble could be
associated with IRDCs (see Fig. \ref{fig8}).
The ammonia emissions provide the approximate $V_{\mathrm{LSR}}$ of these IRDC clumps as presented with
other parameters in Table \ref{t3}.
The fitting and calculation methods in Section \ref{3.2.1}, Section \ref{3.2.2}, and Section \ref{3.3.1} were used to derive these parameters.
Although their radial velocities suggest a spatial relation with the N\,24 bubble system, there is no obvious evidence
of interaction between them and the bubble in the available IR data and molecular emission data.
If these clumps were associated with the bubble, the passing of the expanding bubble would expose the IRDCs to a strong stellar wind and a strong ultraviolet radiation field. This would make these heated IRDCs bright at 8\,$\mu$m. However, they only appear to be infrared dark clouds in the 8\,$\mu$m band. Therefore,
we conclude that these IRDCs are not likely to be directly related to the N\,24 bubble system and are (possibly)
located near the walls of N\,24 in the foreground or background.
Higher resolution data and precise distance measurements are required to further clarify their spatial relation with N\,24.
For the above reasons, these clumps have not (yet) been taken into account as part of the impact of the expanding
bubble on the surrounding regions.

\begin{figure}
\centering
\tiny
\includegraphics[width=0.48\textwidth]{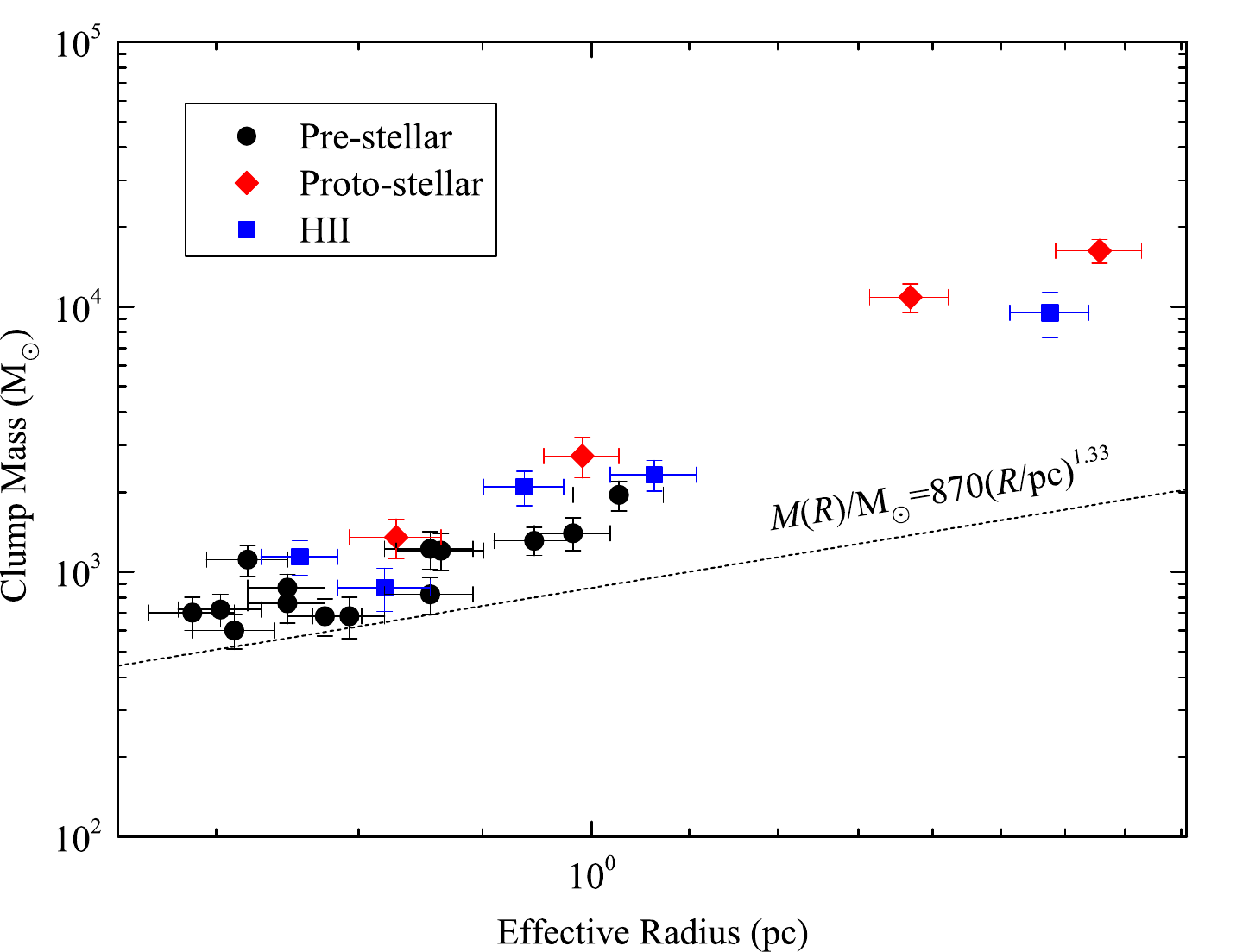}
\caption{Mass versus size for twenty-three clumps at different evolutionary stages.
The black dotted line represents a classic empirical relation of $M(R)/\mathrm{M_{\odot}} = 870 (R/\mathrm{pc})^{1.33}$ serving as a threshold for forming massive stars.}
\label{fig9}
\end{figure}

\begin{figure*}
\centering
\tiny
\includegraphics[width=17cm]{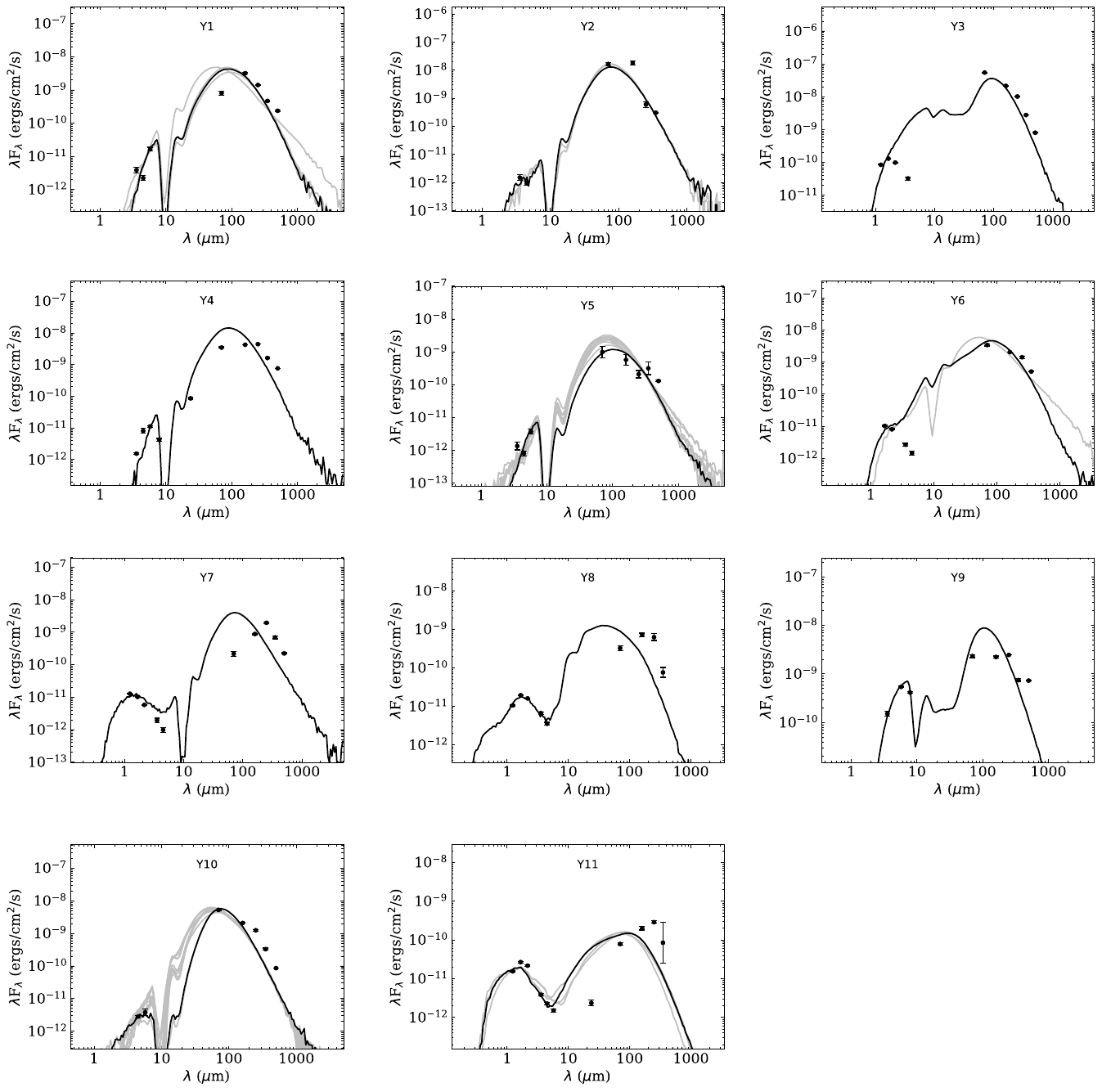}
\caption{SED fitting plots for eleven YSOs. In each panel, the black line shows the best fit, and the grey lines
show subsequent good fits with $\chi^{2}- \chi^{3}_{\mathrm{best}} \leq 3\times N_{\mathrm{data}}$.
The filled circles symbolise the input photometry fluxes.}
\label{fig10}
\end{figure*}

\begin{table*}
  \caption{Parameters of the seven clumps associated with IRDCs inside the bubble.}
  \label{t3}
  \begin{tabular}{cccccccccc}
    \hline\hline
    Number & Clump name & $R_{\mathrm{eff}}$ & $T_{\mathrm{dust, peak}}$ & $N_{\mathrm{H_{2}, peak}}$ & $M_{\mathrm{clump}}$  & Central velocity & Velocity width & $\tau$ &  $T_{\mathrm{kin}}$\\
      &   & (pc) & (K) & (10$^{22}$cm$^{-2}$) & 10$^{3}$M$_{\odot}$ & (km s$^{-1}$) & (km s$^{-1})$ &   &  (K) \\
    \hline
    A&	G\,18.974-00.272&	0.65$\pm$0.03&	18.4$\pm$1.6&	0.37$\pm$0.03&	0.58$\pm$0.08&	66.0$\pm$0.2&	3.0$\pm$0.5&	0.55$\pm$0.62&	             $-$\\
    B&	G\,18.945-00.320&	0.91$\pm$0.04&	19.9$\pm$1.9&	0.30$\pm$0.04&	0.96$\pm$0.14&	         $-$&	        $-$&	          $-$&	             $-$\\
    C&	G\,18.904-00.275&	0.69$\pm$0.03&	18.8$\pm$1.7&	0.46$\pm$0.05&	0.77$\pm$0.11&	64.5$\pm$0.1&	2.7$\pm$0.2&	0.96$\pm$0.28&  	19.1$\pm$1.0\\
    D&	G\,18.836-00.294&	0.99$\pm$0.04&	17.0$\pm$1.4&	0.99$\pm$0.05&	1.96$\pm$0.21&	64.0$\pm$0.1&	2.1$\pm$0.3&	0.98$\pm$0.44&		15.9$\pm$1.3\\
    E&	G\,18.805-00.298&	1.39$\pm$0.06&	15.9$\pm$1.2&	1.11$\pm$0.10&	5.21$\pm$0.60&	65.4$\pm$0.1&	2.3$\pm$0.1&	3.21$\pm$0.21&		14.5$\pm$0.3\\
    F&	G\,18.849-00.377&	1.27$\pm$0.06&	17.1$\pm$1.5&	0.56$\pm$0.08&	3.35$\pm$0.43&	62.3$\pm$0.2&	3.1$\pm$0.4&	          $-$&  	         $-$\\
    G&	G\,18.862-00.419&	0.94$\pm$0.04&	19.3$\pm$1.8&	0.46$\pm$0.06&	1.29$\pm$0.18&	65.5$\pm$0.3&	3.2$\pm$0.3&	2.41$\pm$0.83&	    29.2$\pm$4.5\\
    \hline
  \end{tabular}
 \end{table*}

\begin{table*}
  \caption{Derived parameters of eleven YSOs from the SED best-fit YSO models.}
  \label{t4}
  \begin{tabular}{cccccccccc}
    \hline\hline
    ID & SSTGLMC & $A_{v}$  & Age & $M_{\star}$ & $M_{\mathrm{disk}}$ & $M_{\mathrm{env}}$ & $\dot{M}_{\mathrm{env}}$& Stage & Association \\
     &   & [mag] & [yr] & [M$_{\odot}$] & [M$_{\odot}$] & [M$_{\odot}$] & [M$_{\odot}$yr$^{-1}$] &   &   \\
    \hline
    Y1&  	G019.0549-00.2117&	19.14&	6.10$\times$10$^{4}$&	 9.57&	5.02$\times$10$^{-3}$&	1.52$\times$10$^{3}$&   2.86$\times$10$^{-3}$&	 I&	 Cl.1\\
    Y2&  	G019.0839-00.2768&	25.45&	1.38$\times$10$^{4}$&	14.61&	8.92$\times$10$^{-3}$&	4.56$\times$10$^{2}$&	3.30$\times$10$^{-3}$&	 I&	 Cl.7\\
    Y3&	    G019.0762-00.2873&	 3.90&	3.62$\times$10$^{3}$&	19.75&	                 0.00& 4.10$\times$s10$^{2}$&	7.72$\times$10$^{-3}$&	 I&	 Cl.7\\
    Y4& 	G018.8884-00.4741&	33.98&	3.43$\times$10$^{4}$&	12.51&	1.06$\times$10$^{-1}$& 2.90$\times$s10$^{3}$&	4.60$\times$10$^{-3}$&	 I&	 Cl.9\\
    Y5& 	G018.8921-00.5108&	24.69&	3.46$\times$10$^{4}$&	 8.29&	8.22$\times$10$^{-3}$&	1.44$\times$10$^{3}$&	3.15$\times$10$^{-3}$&	 I&	 Cl.11\\
    Y6& 	G018.8847-00.5095&	 8.60&	2.93$\times$10$^{4}$&	10.05&	8.13$\times$10$^{-3}$&	1.58$\times$10$^{3}$&	2.14$\times$10$^{-3}$&	 I&	 Cl.13\\
    Y7& 	G018.8635-00.4805&	 0.77&	9.13$\times$10$^{3}$&	11.11&	2.73$\times$10$^{-2}$& 1.03$\times$s10$^{2}$&	1.60$\times$10$^{-3}$&	 I&	 Cl.12\\
    Y8& 	G018.8537-00.4834&	 1.96&	3.52$\times$10$^{5}$&	 6.97&	1.81$\times$10$^{-2}$&	2.76$\times$10$^{1}$&	7.59$\times$10$^{-6}$&	II&	 Cl.12\\
    Y9& 	G018.8331-00.4776&	15.30&	1.41$\times$10$^{3}$&	13.69&	                 0.00&  1.40$\times$10$^{3}$&	6.67$\times$10$^{-3}$&	 I&	 Cl.16\\
    Y10&	G018.8251-00.4669&	17.10&	3.95$\times$10$^{3}$&	13.38&	6.59$\times$10$^{-1}$&	1.08$\times$10$^{2}$&	2.21$\times$10$^{-3}$&	 I&	 Cl.16\\
    Y11&	G018.8120-00.4956&	 0.75&	2.25$\times$10$^{4}$&	 2.85&	6.18$\times$10$^{-2}$&	4.92$\times$10$^{2}$&	1.23$\times$10$^{-3}$&	 I&  Cl.16\\
    \hline
  \end{tabular}
 \end{table*}

\subsection{Star formation in clumps}
\subsubsection{YSOs associated with 23 clumps}

The virial ratio and evolutionary phase for these clumps suggest that they have formed stars or have the ability to form them.
For the actual formation of massive stars, the classic empirical mass-size relation of $M(R)/\mathrm{M_{\odot}} = 870 (R/\mathrm{pc})^{1.33}$ was found to be an approximate threshold for massive star formation
\citep{kau10}.
For this reason, the mass versus effective radius relation has been plotted for the 23 clumps in Figure \ref{fig9}
together with the classical relation.
All of our 23 clumps are found to lie above this line and are consistent with being precursors to massive star formation.
In confirmation, \citet{ker13} found intermediate-mass YSOs (~2-10\,M$_{\odot}$) close to most of
the clumps.
However, since \citet{ker13} did not use long-wavelength data, some of the YSOs buried in the clumps may
have gone undetected.

In order to investigate possible YSOs associated with our 23 clumps, eleven point sources at 70\,$\mu$m obtained from \citet{mol16} are
coincident with seven out of the twenty-three clumps, as shown in the left of Figure \ref{fig4}.
The mass range and evolution of these YSOs candidate sources follows from the bolometric luminosity by fitting
the spectral energy distributions (SEDs) with the grid of YSO model SEDs of \citet{rob06}.
This model grid consists of 20,000 two-dimensional Monte Carlo radiation transfer models using linear regression
fits to the multi-wavelength photometry measurements of each given source.
As a first step, the photometry data of each candidate YSO clump
source were extracted from the surveys of 2MASS, $\textit{Spitzer}$-GLIMPSE, $\textit{Spitzer}$-MIPSGAL
and $\textit{Herschel}$-HiGAL archives from the IRSA data base\footnote{\url{http://irsa.ipac.caltech.edu/frontpage/}} \citep[i.e.,][]{liu18, das16}.
As a second step, a visual extinction in the range of 0-40 mag was estimated from the
column density map by a relation of $A_{v}=5.34\times10^{-22}\textit{N}_{\mathrm{H}_{2}}$, assuming an average (kinematic) distance of 4.3-4.7\,kpc for N\,24.
The optimum SEDs within a specific $\chi^{2}$ were returned based on a regression algorithm as presented in
Figure \ref{fig10}, where all other models that satisfy $\chi^{2}- \chi^{3}_{\mathrm{best}} \leq 3\times N_{\mathrm{data}}$
are also shown with $N_{\mathrm{data}}$ being the number of data points.
Several calculated key parameters of the fits for the eleven candidate sources are tabulated in Table \ref{t4}.
According to these parameters, \citet{rob06} classified YSOs into three stages:
\newline Stage I has $\dot{M}_{\mathrm{env}}/M_\star>10^{-6}$yr$^{-1}$,
\newline Stage II has $\dot{M}_{\mathrm{env}}/M_\star<10^{-6}$yr$^{-1}$ and $M_{\mathrm{disk}}/M_\star>10^{-6}$, and
\newline Stage III has $\dot{M}_{\mathrm{env}}/M_\star<10^{-6}$yr$^{-1}$ and $M_{\mathrm{disk}}/M_\star<10^{-6}$,
\newline where $\dot{M}_{\mathrm{env}}$ is the envelope accretion rate, $M_\star$ is the stellar mass, and $M_{\mathrm{disk}}$
is the disk mass.
Most of the best-fit model masses are larger than 8\,M$_{\odot}$, while the evolution stage is almost uniformly at
Stage I.
Although two YSOs (Y8 \& Y11) have a smaller mass, they are all clumps that can form massive stars,
which is consistent with our previous evaluation.

\subsubsection{Outflow of G\,19.07-00.28}

\begin{figure}
\centering
\tiny
\includegraphics[width=0.48\textwidth]{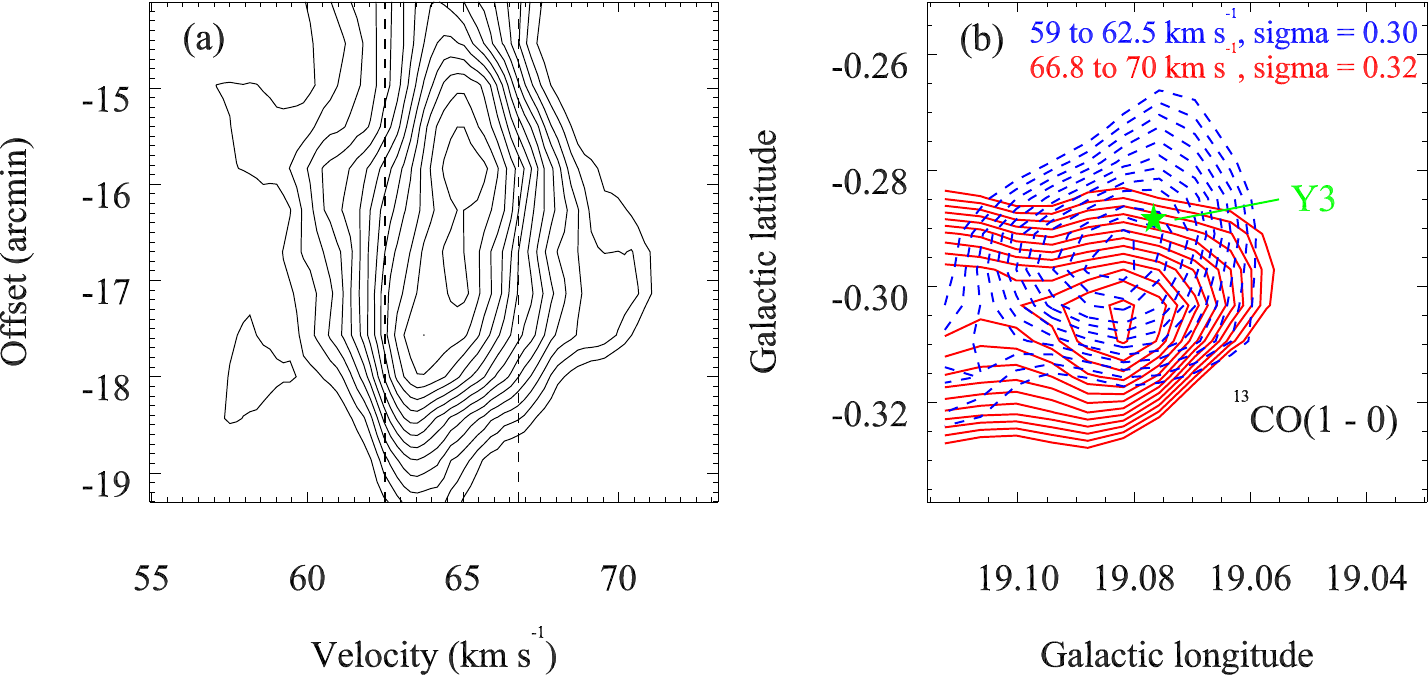}
\caption{The outflow in G\,19.07-0.28 and YSO candidate Y\,3.
(a) The position-velocity (P-V) slice diagram. Contour levels are from 0.6 to 3.8 \,K by 0.267\,K.
(b) Contour maps of the wing emissions integrated over 59.0 to 62.5\,km s$^{-1}$ for the blue wing (blue
dashed line) and 66.8 to 70.0\,km s$^{-1}$ for the red wing (red solid line), respectively.
Both blue and red contours start from 10$\sigma$ and increase by 0.5\,$\sigma$.
The green pentagram indicates the position of YSO candidate Y3.}
\label{fig11}
\end{figure}

The $^{13}$CO emission in the region may be used to find outflows from these star forming regions.
In the G\,19.07-0.28 region where Y\,3 is located, we surprisingly find an apparent outflow in the line wing of
the GRS\,$^{13}$CO\,(1-0) profile, but there are no obvious signs of outflows in other regions.
Figure \ref{fig11}(a) shows the position-velocity (PV) diagram of G\,19.07-0.28 region, which determined the
velocity range of the red and blue lobes.  The $^{13}$CO\,(1-0) integrated intensity images of the outflow lobes
in Figure \ref{fig11}(b) shows the spatial extent of the redshifted and blueshifted lobes.
Assuming that the $^{13}$CO emission is optically thin and that all levels have the same excitation temperature, we
used the equation derived in \citet{gar91} to estimate the total column density:
     \begin{equation}\label{eq-9}
     N_{\mathrm{tot}}(\mathrm{^{13}CO}) = 4.56 \times 10^{13} \frac{\textit{T}_{\mathrm{ex}}+0.88}{\mathrm{exp}({-5.29/\textit{T}_{\mathrm{ex}}})}\int \frac{\textit{T}_{\mathrm{a}}^{*}}{\eta_{\mathrm{mb}}},
     \end{equation}
where $T_{\mathrm{a}}^{*}$ is the observed antenna temperature of $^{13}$CO, and $\eta_{\mathrm{mb}}$ is the main
beam efficiency of 0.48. The excitation temperature $T_{\mathrm{ex}}$ is assumed to be 30\,K for the high-mass sources
\citep{she96a, she96b, beu02, wu04, wu05, xu06}.
The column density of H$_{2}$ gas was required to estimate the outflow masses and a conversion factor
$\mathrm{[H_{2}/^{13}CO]}=5.0\times10^{5}$ was assumed for $^{13}$CO lines.
The mass for each outflow lobe was derived via the formula:
      \begin{equation}\label{eq-10}
      M_{\mathrm{r/b}}= \sum N_{\mathrm{tot}}(\mathrm{^{13}CO})[[\mathrm{H_{2}/^{13}CO}]\mu_{\mathrm{H_{2}}}\textit{m}_{\mathrm{H}}\textit{A}_{\mathrm{pixel}},
      \end{equation}
where $\mu = 2.8$ is the mean molecular weight, $m_{\mathrm{H}}$ is the mass of an hydrogen atom, and
$A_{\mathrm{pixel}}$ is the area of each pixel within the outflow lobe defined by the lowest (3$\sigma$) contours.
The mass of each lobe was obtained by summing over all spatial pixels and the total outflow mass was
obtained by adding the masses of each lobe, $M_{\mathrm{out}} = M_{\mathrm{r}} + M_{\mathrm{b}}$.
Subsequently, the momentum and kinetic energy within a velocity channel were calculated for each lobe by:
      \begin{equation}\label{eq-11}
      P_{v, \mathrm{r/b}}= \sum M_{v, \mathrm{pixel}} \times v
       \end{equation}
      \begin{equation}\label{eq-12}
      E_{v, \mathrm{r/b}}= \sum \frac{1}{2}M_{v, \mathrm{pixel}} \times v^{2}
       \end{equation}
where \textit{v} is the velocity of each channel relative to the systemic velocity, and $M_{v, \mathrm{pixel}}$ is the
mass for each pixel corresponding to the emission in that channel.
The velocity range for the red and blue wings of $^{13}$CO spectra were estimated to be $\Delta v_{\mathrm{r}}=[66.8, 70]$
(km s$^{-1}$) and $\Delta v_{\mathrm{b}}=[59, 62.5]$ (km s$^{-1}$).
The total momentum and kinetic energy were obtained in a similar manner as for the total outflow mass.
The dynamical timescale $t_{\mathrm{dyn}}$ of the outflow is calculated as: $t_{\mathrm{dyn}}=l/v$, where \textit{l} is the
separation between the peaks of the red and blue lobes and \textit{v} is the mean outflow velocity obtained by
$P_{\mathrm{out}}/M_{\mathrm{out}}$.
The mass rate of the outflow, the mechanical force, and the mechanical luminosity of the molecular outflow
are derived as:
      \begin{equation}\label{eq-13}
      \dot{M}_{\mathrm{out}}= M_{\mathrm{out}} / t_{\mathrm{dyn}},
      \end{equation}
     \begin{equation}
 \label{eq-14}
        F_{\mathrm{out}}= P_{\mathrm{out}} / t_{\mathrm{dyn}},
                      \end{equation}
      \begin{equation}
\label{eq-15}
       L_{\mathrm{out}}= E_{\mathrm{out}} / t_{\mathrm{dyn}},
              \end{equation}
Finally, we presented the computed results of $M_{\mathrm{r}}$, $M_{\mathrm{b}}$, $M_{\mathrm{out}}$, $P_{\mathrm{out}}$,
$E_{\mathrm{out}}$, $t_{\mathrm{dyn}}$, $\dot{M}_{\mathrm{out}}$, $F_{\mathrm{out}}$and $L_{\mathrm{out}}$ as 291.01 $\pm$ 56.98 M$_{\odot}$,
214.50 $\pm$ 42.50 M$_{\odot}$, 505.51 $\pm$ 99.48 M$_{\odot}$, 1825.60 $\pm$ 331.81 M$_{\odot}$km s$^{-1}$, 71.35 ( $\pm$ 14.04) $\times$ 10$^{45}$ ergs,
13.00 ( $\pm$ 3.48) $\times$ 10$^{4}$ yr, 38.35 ( $\pm$ 12.74) $\times$ 10$^{-4}$ M$_{\odot}$yr$^{-1}$, 138.51 ( $\pm$ 44.82) $\times$ 10$^{-4}$
M$_{\odot}$km s$^{-1}$yr$^{-1}$, and 4.54 $\pm$ 1.51 L$_{\odot}$, respectively.
These values are consistent with the mean value for a large sample of massive star-forming regions \citep{li18}.

\subsection{Triggered star formation scenario}\label{sec 3.5}

In order to find out whether fragmentation can really be produced by the collected material, we estimated the
fragmentation time of the collected layer following \citep{whi94b}:
       \begin{equation}\label{eq-16}
       t_{\mathrm{frag}}=1.56(\frac{\alpha_{\mathrm{s}}}{0.2})^{7/11}(\frac{N_{\mathrm{Lyc}}}{10^{49}})^{-1/11}(\frac{n_{0}}{10^{3}})^{-5/11},
       \end{equation}
where $\alpha_{\mathrm{s}}$ is the turbulent velocity, $N_{\mathrm{Lyc}}$ is the Lyman continuum photon flux, and n$_{0}$
is the original ambient density.
The $N_{\mathrm{Lyc}}$ can be estimated following \citep{kur94}:
       \begin{equation}\label{eq-17}
       N_{\mathrm{Lyc}}=7.588\times10^{48}(\frac{T_{\mathrm{e}}}{\mathrm{K}})^{-0.5}(\frac{\nu}{\mathrm{GHz}})^{0.1}(\frac{S_{\nu}}{\mathrm{Jy}})(\frac{D}{\mathrm{kpc}})^{2},
       \end{equation}
and using $\nu=1.4$ GHz, $T_{\mathrm{e}}=7500$ K, $S_{\nu}=40$ Jy and $D=4.5$ kpc given by \citet{ker13},
we find value of $N_{\mathrm{Lyc}}=7.34\times10^{49}$ photons s$^{-1}$.
To estimate $n_{0}$, the total mass distributed within the shell was taken as the sum of the masses of the
23 detected clumps (at the right of Figure \ref{fig4}), which is $\sim$ 63000\,M$_{\odot}$. Added to this should be
the mass of the ionised component ($\sim$ 7000\,M$_{\odot}$), which was estimated via
$M_{\mathrm{ion}} = 4/3 \pi r^{3}_{\mathrm{HII}}n_{\mathrm{e}}m_{\mathrm{p}}$, where $r_{\mathrm{HII}}$ is taken to be
the radius of the bubble, $n_{\mathrm{e}}$ is the electron density of $\sim$ 20\,cm$^{-3}$ \citep{ker13}, and $m_{\mathrm{p}}$
is the proton mass.
Assuming that N\,24 has a spherical structure, we found $n_{0} = 2.01 \times 10^{2}$\,cm$^{-3}$.
Using the typically adopted range for the turbulent velocity $\alpha_{\mathrm{s}}$ = 0.2-0.6\,km s$^{-1}$,
we then obtained a fragmentation time range $t_{\mathrm{frag}} = 2.70-5.44$\,Myr.
It should be noted that the stellar winds have not been taken into account in the estimate of the fragmentation time. Therefore, the time derived from Equation (\ref{eq-17}) will be an overestimate. Given this uncertainty, the fragmentation time-scale is considered comparable to the dynamical age of $t_{\mathrm{dyn}}$ = $\sim$ 1.5-4.0 Myr \citep{ker13} of the HII region. This suggests that the collected molecular cloud has had enough time to fragment into molecular cores during the lifetime of N\,24. In addition, the spatial distribution of YSOs \citep[see Figure 10 in][]{ker13} and clumps show an overabundance on the bubble rim relative to the surrounding ISM. Combined with the existence of ionization compression and an overabundance of YSOs and clumps, the time-scales demonstrate that the C \& C mechanism is a possible scenario for the star formation at the periphery of N\,24. However, since the IRDC clumps exist on the SW edge of the bubble (see Fig. \ref{fig4} (right)), we cannot ignore the possibility that the RDI mechanism may have been at work there.

\section{Summary}\label{sec4}

We have presented a multi-wavelength investigation towards the large Galactic IR bubble N\,24 to analyse the physical
properties of the dust and gas therein and to explore the possibility of triggered star formation by combining archival survey data with our molecular line observations by NSRT. The main results of our study are summarised as follows:

(1) The infrared structure and the distribution of the molecular emissions show that the two main regions of
G\,19.07-0.28 and G\,18.88-0.49 in the N\,24 shell are consistent with star formation triggered by the expanding bubble.
As a result of the feedback from massive stars, some new bubbles have already formed in these two regions,
which further affect the environs therein.

(2) Several filamentary structures of IRDCs appear in absorption at 8\,$\mu$m and are clearly visible in the
far-infrared images from \textit{Herschel} with widespread wavelength coverage and at high resolution.
We estimated the $V_{\mathrm{LSR}}$ of the IRDCs inside the bubble in order to determine the association of these clumps, but we
could not accurately determine their spatial correlation with the N\,24 bubble from the available data.

(3) There is an anti-correlation between the distribution of the dust temperature and the column density in clumps,
which may be attributed to a lower penetration of the external heating from the H\,{\sc{ii}} region into the dense regions.

(4) We found 23 dense dust clumps in the column density distribution map using the Clumpfind2d algorithm, which are
almost all distributed along the bubble shell.
These clumps have a mean size of ~0.92 $\pm$ 0.06\,pc, a mean peak temperature of ~20.8 $\pm$ 0.5\,K, a mean column density of
~0.86 ( $\pm$ 0.19) $\times$ 10$^{22}$ cm$^{-2}$, a mean mass of ~2.66 ( $\pm$ 0.81) $\times$10$^{3}$ M$_{\odot}$, and a mean volume density of
~7.75 ( $\pm$ 0.46) $\times$ 10$^{3}$ cm$^{-3}$.
The column density and volume density of clumps affected by second generation bubbles are larger than for
others, indicating that these second generation bubbles are also expanding and sweeping up material.

(5) We classified the 23 dust clumps according to their evolutionary stage.
The values of the virial ratio of clumps in a proto-stellar stage and most of those in the H\,{\sc{ii}} stage are
significantly larger than unity, which indicates that these are likely to be gravitationally bound systems and
will collapse to form stars.

(6) The mass-size distribution of the dense clumps suggests that they all could form massive stars. The SED fitting results of the eleven identified YSOs indicate that nine of these have a mass above 8\,M$_{\odot}$.

(7) The G\,19.07-0.28 region is found to contain an H\,{\sc{ii}} region and a massive star Y\,3 and the parameters of the apparent outflow have been derived from the line wings of the GRS\,$^{13}$CO\,(1-0) emission.

(8) The compatibility of the dynamical and fragmentation time-scales and the overabundance of YSOs and clumps on the rim suggest that the C \& C mechanism is in play at the boundary of the bubble, but the co-existence of the IRDC at the edge of the bubble indicates that RDI may also play a role here.

\section*{Acknowledgements}

This work has been funded by The National Natural Science foundation of China under grants Nos. 11433008,
11703074, 11603063, and the CAS "Light of West China" Program under Grant 2018-XBQNXZ-B-024, 2016-QNXZ-B-23, and The Heavenly Lake Hundred-Talent Program of Xinjiang Uygur Autonomous Region of China. This work is based on observations made with the Nanshan 26 meter radio telescope, which is operated by the Key Laboratory of Radio Astronomy, Chinese Academy of Sciences. WAB has been funded by Chinese Academy of Sciences President's International Fellowship Initiative. Grant No. 2019VMA0040.

\end{document}